%% file: njp_ms_pot_V3.tex
\documentclass[12pt]{iopart}

\usepackage{graphicx,color,caption,subfigure}
\usepackage{amsfonts,amssymb,amsthm,latexsym}
\expandafter\let\csname equation*\endcsname\relax
\expandafter\let\csname endequation*\endcsname\relax
\usepackage{amsmath}

\theoremstyle{definition}

\newcommand{\cL}{\mathcal L}

\newcommand{\eps}{\epsilon}

\newcommand{\cM}{{\mathcal M}}

\newcommand{\R}{{\mathbb R}}

\renewcommand{\S}{{\mathbb S}}
\newcommand{\T}{{\mathbb T}}
\newcommand{\LL}{{\mathcal L}}

\begin{document}
 \setlength{\baselineskip}{15pt}

\title{Noise-induced transitions in rugged energy landscapes}

\author{A Duncan$^1$, S Kalliadasis$^2$, G A Pavliotis$^1$, M Pradas$^3$}
\address{$^1$ Department of Mathematics, Imperial College London, London SW7 2AZ, UK }
\address{$^2$ Department of Chemical Engineering, Imperial College London, London SW7 2AZ, UK}
\address{$^3$ Department of Mathematics and Statistics, The Open University, Milton Keynes MK7 6AA, UK}


\ead{}
\vspace{10pt}

\begin{abstract}
We consider the problem of an overdamped Brownian particle moving in
multiscale potential with $N+1$ characteristic length scales: the macroscale
and $N$ separated microscales. We show that the coarse-grained dynamics is
given by an overdamped Langevin equation with respect to the free energy and
with a space dependent diffusion tensor, the calculation of which requires
the solution of $N$ fully coupled Poisson equations. We study in detail the
structure of the bifurcation diagram for one-dimensional problems and we show
that the mulitscale structure in the potential leads to hysteresis effects
and to noise-induced transitions. Furthermore, we obtain an explicit formula
for the effective diffusion coefficient for a self-similar separable
potential and we investigate the limit of infinitely many small scales.

\end{abstract}

%
%
\submitto{\NJP}
%
%
%

\section{Introduction}
\label{sec:intro}

Brownian motion in disordered media (or rugged energy landscapes) is a
problem of great scientific and technological interest, and applications are
found in a wide range of different areas, such as e.g.~collective transport
of particles in random media
\cite{duncan2015,DEPS2015,PavSt05b,HP07,PavlVog08,Pap95},  molecular
motors~\cite{LatPavlKram2013,pavl05}, and protein reaction dynamics and
folding~\cite{pollak2008}, to name but a few. In the latter example in
particular, proteins are dynamic macromolecules that exhibit many scales of
molecular motion which is governed by a hopping mechanism through the local
minima of the free-energy surface, the so-called conformational substrates or
microstates. Understanding the effect of the microstates on the large scale
dynamics of proteins is a problem of both theoretical and practical interest.
At the same time, a rugged energy landscape can introduce metastability in
the system~\cite{Bovier_al_2004,Bovier_2015} and the degree of metastability
can increase with the complexity of the landscape, invalidating predictions
based on thermodynamic arguments, e.g.~\cite{SavKalPavl10}. In addition,
other systems characterised by the presence of rugged energy landscapes
include flows in structured or disordered media such as fluid flow in porous
media~\cite{Pavliotis-al12,SPPK2013,SPK2014} or contact line dynamics on
chemically and/or topographically heterogeneous
substrates~\cite{SavKalPavl10,SavKalPavl10a,SavKalPavl10b,Vellingiri2011,Wylock2012};
while the understanding of conformational changes in complicated multiscale
energy landscapes can have significant impact to technological applications
such as crystallisation~\cite{Threlfall2010} and drug
design~\cite{Mortier_al2015}.

The dynamics of a Brownian particle moving in a rugged energy landscape can
be modeled using the Langevin dynamics, either non-Markovian~\cite[Ch.
8]{Pavl2014}, underdamped or the overdamped (Smoluchowski)
dynamics in a multiscale potential which 
can be taken to be either deterministic or random. The main goal of the
present work is to study in detail the coarse-grained dynamics of the
Smoluchowski dynamics in an $N$-scale periodic potential. In particular, we
will derive rigorously the coarse-grained dynamics and study the quantitative
and qualitative properties of the homogenized model. It is important to note
that many interesting phenomena, such as subdiffusion, may arise in the
coarse grained dynamics systems with a multiscale potential, and, as it was
shown in \cite{Zwan88}, the presence of a microscale (``roughness") in the
potential decreases the mean first passage time.  In particular, given a potential $V(x)$ with two metastable states, perturbing $V(x)$ with microscale fluctuations would result in a decrease in the escape/reaction rate between the two states, see also \cite{HanTalkBork90}.  Such
results can be obtained in a systematic and rigorous way using analytical
multiscale techniques.  One can also approximate the effective dynamics numerically, using methods such as heterogeneous multiscale methods~\cite{abdulle2014fully}, reduced basis finite element heterogeneous multiscale methods, \cite{abdulle2012reduced}, as well as equation-free methods~\cite{Gear2003a,Kevrekidis2009}. 
%

Here, we further assume an overdamped Langevin dynamics of a Brownian
particle moving in a multiscale periodic potential, where the macroscale is
assumed to be confining (see  figure \ref{Fig: potentials} for some examples
of multiscale potentials). By carefully analysing the corresponding effective
(averaged) equation in different examples of potentials we are able to
observe nontrivial dynamics which emerges as a consequence of the interplay
between noise level and microscopic structure. In particular, we find that
the microscopic fluctuations conspire with the additive noise to produce
noise-induced hysteresis and noise-induced stabilisation depending on the
particular choice of the potential. In all cases, we are able to fully
characterise the different state transitions in terms of critical exponents.

%
%
Our basic model will be the first-order Langevin equation
\begin{equation}
d X^\eps_t = - \nabla V_{\eps} \left(X^\eps_t \right)  dt + \sqrt{2 \sigma } d W_t,
\label{e:main}
\end{equation}
where $W_t$ denotes standard Brownian motion on $\R^d$ and where the magnitude of the variance of the noise $\sigma$ would typically be related to the inverse temperature. The potential depends on $N+1$ scale, the macroscale and $N$ small scales:
\begin{equation}\label{e:potential}
V_{\eps}(x) = V \left(x, \frac{x}{\eps}, \frac{x}{\eps^{2}}, \dots \frac{x}{\eps^{N}} \right),
\end{equation}
and it  is assumed to be confining at the macroscale and periodic in all small scales (detailed assumptions on the potential will be presented in the next section). 
For the dynamics~\eqref{e:main}, with the potential~(\ref{e:potential}) tools
from homogenization theory, in particular reiterated
homogenization~\cite{AllBri96} can be used in order to obtain an effective
equation, valid in the limit of infinite scale separation $\eps \rightarrow
0$.  

Several aspects of this problem have already been studied. First, for
periodic potentials with one characteristic length scale, under the diffusive
rescaling $X^{\eps}_t := \eps X_{t/\eps^2}$ the effective diffusive dynamics
becomes diffusive with an effective diffusion matrix $D$ that can be
calculated by solving an appropriate Poisson equation, posed on the unit
periodicity cell~\cite[Ch. 13]{PavlSt08}, \cite[Sec. 3.4]{lions}. This result
is a form of the functional central limit theorem for diffusion processes
with periodic coefficients~\cite{bhatta}. Furthermore, diffusion is always
depleted and it becomes exponentially small in the limit
$\sigma \rightarrow 0$~\cite{Kozlov89a}. The case of Brownian dynamics
in a two-scale separable potential was studied in~\cite{PavlSt06}. In
particular, the dynamics~\eqref{e:main} with a potential $V$
in~\eqref{e:potential} of the form $V ( x, y ; \alpha) = \alpha V(x) + p(y)$,
with $p(\cdot)$ a smooth periodic function, was considered. It was shown
in~\cite{PavlSt06} that the maximum likelihood estimator for the coefficients
in the drift of the homogenized equation, given observations from the full
dynamics~\eqref{e:main}, is asymptotically biased.

On the other hand, the problem of homogenization for Brownian particles in
periodic potentials with $N-$scales, in the absence of a
macroscopic/confining potential was studied in~\cite{BenArOw03, Owhadi2003}.
In these papers the overdamped Langevin dynamics in potentials of the form
\begin{equation}\label{e:arous}
V^N(x) = \sum_{k=1}^N U_k \left(\frac{x}{R_k} \right),
\end{equation}
where  $U_k, \, k=1,\dots$ are H\"{o}lder continuous periodic potentials.
Under the assumption that the scale ratios $\frac{R_{k+1}}{R_k}$ are bounded
from above and below, i.e. when no scale separation is present, it was shown
that the eigenvalues of the effective diffusivity tensor $D(V^N)$  decay exponentially quickly as
the number of scales increases. Using this result, the authors were able to
show that in the limit of infinitely many scales the effective behavior is
characterized by anomalous slow behavior. This subdiffusive behavior can be
analyzed in a quantitative way by studying the mean exit time of the
effective dynamics from a ball whose radius is of $O(1)$.

The potential~(\ref{e:potential}) that we consider here can be thought of as
a caricature of a disordered medium.
For self-similar potentials of the form
\begin{equation}\label{e:pot-separable}
V_{\eps}(x) = \sum_{j=1}^{+\infty} V \left( \frac{x}{\eps^{j}} \right),
\end{equation}
where $V(\cdot)$ is a periodic function, it is possible, at least in one
dimension, to obtain an analytical formula for the effective diffusion
coefficient.

Here we will show that the coarse-grained equation of (\ref{e:main}) is reversible with respect to an appropriate Gibbs
measure and that the effective potential is given by a coarse-grained free energy. In addition, an important point to note is that, 
even though the noise in the full dynamics~\eqref{e:main} is additive (since
it is due to thermal fluctuations) the noise in the coarse-grained model is
multiplicative. It is well-known that multiplicative noise can lead to
noise-induced state transitions, both first- and
second-order~\cite{MackeyLongtinLasota1990}. The fact that additive noise
from the fast scales, combined with the multiscale nature of the dynamics,
leads to multiplicative noise and noise-induced transitions in the
coarse-grained dynamics, was shown rigorously and investigated in detail for
the stochastic Kuramoto-Shivashinsky (sKS) equation--an SPDE with no gradient
structure~\cite{Pradas2012, Pradas2011}. Specifically, as was shown in these
studies, the coarse-grained dynamics of the sKS equation near the instability
threshold is described by a low-dimensional stochastic differential equation
(SDE) (an ``amplitude equation") of the Landau-Stuart type with additive as
well as multiplicative noise. For particular choices of the spatial
correlation structure of the noise, the amplitude equation contains only
multiplicative noise that leads to noise-induced stabilization and
intermittent behavior. The transition between the three possible states of the
system--normal, Gaussian-like behavior, intermittency and
stabilization--depends on the strength of the noise.

One of our goals here is to investigate similar issues for the multiscale
overdamped Langevin dynamics. In particular, following the techniques
developed in~\cite{MackeyLongtinLasota1990}, see also~\cite[Sec.
5.4]{Pavl2014}, for non-multiscale SDEs with multiplicative noise in one
dimension, we analyze the effect of the multiscale structure on the
bifurcation diagram of the coarse-grained dynamics in one dimension. In
particular, we show that the presence of several spatial scales leads to
hysteresis loops in the bifurcation diagram  which
we can characterize quantitatively in terms of an appropriate critical
exponent. We note the similarities between our findings and the work on
critical transitions and bifurcation theory for non-autonomous stochastic
dynamical systems, in particular the emergence of hysteresis phenomena in the
study of the so-called tipping points~\cite{Kuehn2011}.  A similar numerical study of water molecules filling or emptying carbon nanotubes was investigated in \cite{sriraman2005coarse}, where a coarse grained potential energy landscape was derived computationally, using coarse-grained molecular dynamics, and use to investigate the metastability and hysteretic parameter dependence of the dynamics.

The rest of the paper is organized as follows: In Section~\ref{sec:homog} we
present the model that we will be considering in detail and we also give our
main results: the formula for the homogenized equation and the main
properties of the effective potential (free energy) and of the effective
diffusion tensor. The effect of the multiscale structure of the potential on
a pitchfork bifurcation is studied in Section~\ref{sec:pitchfork}.
Noise-induced stabilization phenomena for multiscale potentials are
considered in Section~\ref{subsec:stabiliz}. In Section~\ref{sec:piecewise}
we calculate the effective diffusion coefficient for a Brownian particle
moving in a piecewise linear self-similar potential with infinitely many
scales. Conclusions and a discussion are offered in
Section~\ref{sec:conclusions} and the derivation of the coarse-grained
equation and the calculation of the effective diffusion coefficient using
multiscale techniques are outlined in the Appendices.
%
%

\section{Brownian motion in a rugged energy landscape}
\label{sec:homog}

\begin{figure}
\includegraphics[width=1.0\textwidth]{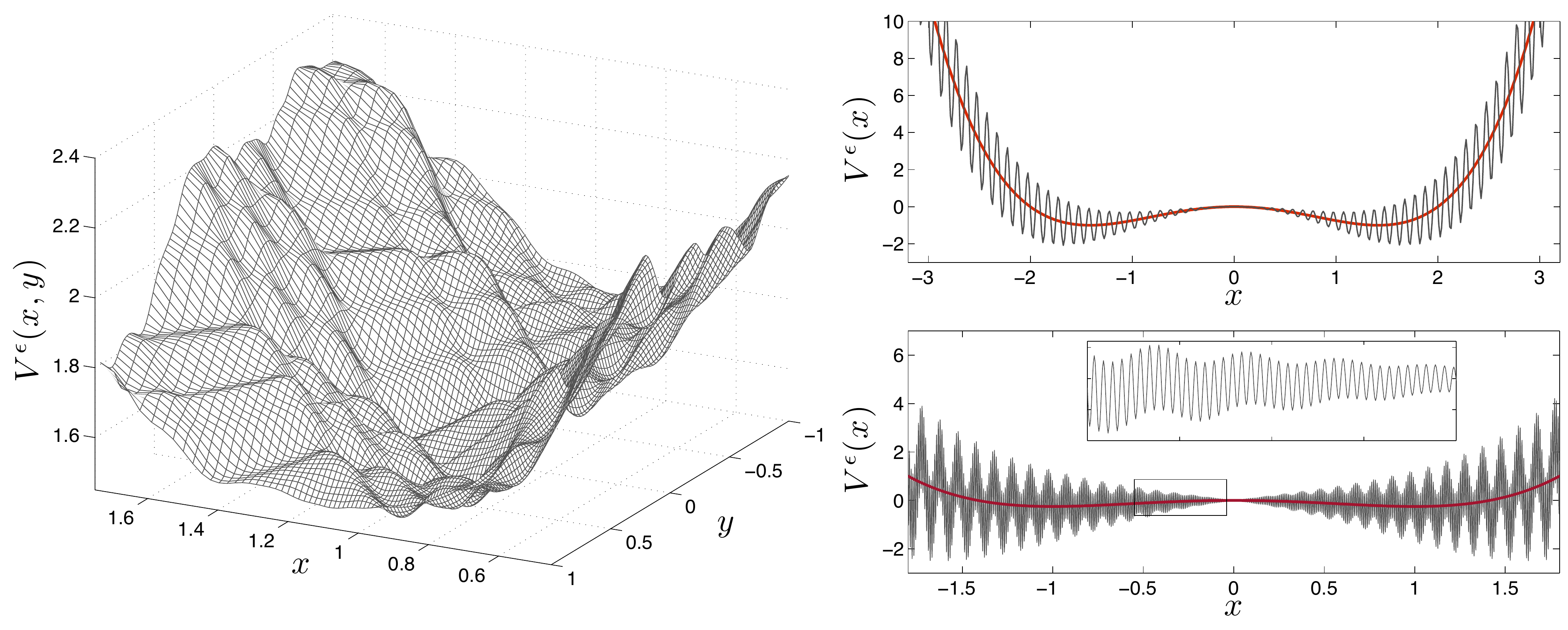}
\centering
\caption{Examples of multiscale potentials. The left panel depicts a two-dimensional rough surface which corresponds to the interfacial interfacial energy of a
droplet on a chemical heterogeneous substrate (see
e.g.~\cite{Vellingiri2011}). The right panels correspond to the case of a one-dimensional (1D)
periodic multiscale potential given by  (\ref{eq:pitchfork_potential}) with
$\alpha=0.5$ (top) and  (\ref{Eq:3s pot}) (bottom). The inlet of the bottom
panel is a zoom into the area marked by a rectangle.} \label{Fig:
potentials}
\end{figure}

We consider the overdamped Langevin dynamics in a multiscale potential with $N+1$ characteristic length scales. The dynamics is given by the following SDE:
\begin{equation}
\label{e:gen-sde}
d X^\epsilon(t) = -\nabla V^\epsilon\left(X^\eps (t)\right)dt+\sqrt{2 \sigma} dW(t),
\end{equation}
where the potential $V^\epsilon(x)$ is of the form
\begin{equation}\label{e:pote-form}
V^\epsilon(x) = V\left(x,\frac{x}{\epsilon},\frac{x}{\epsilon^2},\dots,\frac{x}{\epsilon^N}\right),
\end{equation}
where $\epsilon \ll 1$ measures the degree of scale separation and where
$V(x, y_1, \ldots, y_N)$ is a smooth function which is periodic in all but
the first variable. The variables $y_1, \ldots, y_N$ characterise the
microscopic scales of the potential while $x$ represents the macroscale. So
$V$ is assumed to have a fractal-like structure which is realistic and allows
for analytical progress to be made. Also, without loss of generality, we may
also assume that $V$ has period one in each microscopic variable. $W(t)$
denotes standard Brownian motion in $\R^d$ and $\sigma > 0$ corresponds to
the temperature.  We shall assume that the potential can be decomposed as follows:
\begin{equation}
\label{e:pot-mean}
	V(x, y_1, \ldots, y_N) = V_0(x) + V_1(x,y_1, \ldots, y_N),
\end{equation}
where $V_0$ is assumed to be confining potential, while $V_1$ is assumed to be bounded uniformly with respect to all parameters, and periodic with period $1$ with respect to the variables $y_1,\ldots, y_N$.  This ensures that both the full dynamics~\eqref{e:gen-sde} and
the coarse-grained dynamics,  ~\eqref{e:homog} below, are ergodic~\cite[Sec.
4.5]{Pavl2014}. In particular, the process $\{X^\eps(t) \}$ is
(exponentially) ergodic\footnote{It converges exponentially fast to the
invariant distribution. Details about the rigorous study of \eqref{e:gen-sde} can be found in~\cite{DuncanPavl2016}}
with invariant distribution
\begin{equation}
\rho^\eps(x) = \frac{1}{Z^\eps} e^{-V\left(x,\frac{x}{\epsilon},\frac{x}{\epsilon^2},\dots,\frac{x}{\epsilon^N}\right)/\sigma}, \quad Z^\eps := \int_{\R^d}
e^{-V\left(x,\frac{x}{\epsilon},\frac{x}{\epsilon^2},\dots,\frac{x}{\epsilon^N}\right)/\sigma} \, dx.
\label{e:xeps_inv_meas_ddim}
\end{equation}
The dynamics $\{X^\eps(t) \}$ given by~\eqref{e:gen-sde} is reversible with
respect to the distribution~\eqref{e:xeps_inv_meas_ddim}. In particular, the
generator of the process $\{X^\eps(t) \}$ is self-adjoint in the space
$L^2(\R^d ; \rho^{\eps}(x))$ and can be written in the form
\begin{equation}\label{e:gen-selfadj}
\cL^{\eps} \cdot = \frac{\sigma}{\rho^{\eps}(x)} \nabla \cdot \big(\rho^{\eps}(x) \nabla \cdot \big).
\end{equation}
Introducing the auxiliary variables $y_n = \frac{x}{\eps^n}, \; n=1, \dots N$
and using the chain rule we can write~\eqref{e:gen-sde} as a system of
interacting diffusions across scales:
\numparts
\begin{eqnarray*}
d X^\eps(t) &=& -\nabla_{x} V(X^\eps (t), Y^\eps_1(t), \dots , Y_N^{\eps}(t)) \, dt \nonumber \\
&& - \sum_{\ell =1}^N\frac{1}{\eps^{\ell}}\nabla_{y_{\ell}} V(X^\eps (t), Y^\eps_1(t), \dots , Y_N^{\eps}(t)) \, dt
 + \sqrt{2 \sigma} \, dW(t), \  \label{e:x} \\
d Y^\eps_n(t) &=& -\frac{1}{\eps^n}\nabla_{x} V(X^\eps (t), Y_1(t), \dots , Y_N^{\eps}(t)) \, dt \nonumber \\
&&- \sum_{\ell =1}^N\frac{1}{\eps^{n+\ell}}\nabla_{y_{\ell}} V(X^\eps (t), Y^\eps_1(t), \dots , Y_N^{\eps}(t)) \, dt 
 + \sqrt{\frac{2 \sigma}{\eps^{2n}}} \, dW(t),  \ \label{e:y}
\end{eqnarray*}
\endnumparts
for $ n = 1,\dots N$. The state space of the diffusion process $\{X^\eps(t), \, Y^\eps_1(t), \dots
Y^\eps_N(t)  \}$ is $\R^d \times \T^d \times \dots \times\T^d $ where $\T^d$
denotes the unit torus. This auxiliary diffusion process inherits
from~\eqref{e:gen-sde} the properties of ergodicity and reversibility. Our
goal is to eliminate the fast scales $\{ Y^\eps_1(t), \dots Y^\eps_N(t)  \}$
and to obtain a closed equation for the macroscopic variable $X(t)$. We
remark on the similarity between this homogenization/coarse-graining problem
and that of the derivation of a mean-field limit equation for interacting
diffusions~\cite{Dawson1983}. In \ref{app:Reiterated} we use homogenization
theory~\cite{PavlSt08} and in particular the theory of reiterated
homogenization~\cite{AllBri96} to derive such a closed, SDE for the
macroscopic variable $X^{\eps}(t)$, valid in the limit of infinite scale
separation $\eps \rightarrow 0$. In this section we present the
coarse-grained model and we elucidate some of its main properties.  In particular, in \ref{app:Reiterated} we derive the following result: the
solution $X^{\eps}_{t}$ of~\eqref{e:gen-sde} converges as $\eps
\rightarrow 0$ to the solution of the SDE
\begin{equation}\label{e:homog}
d X_{t} = -\cM(X_{t}) \nabla \Psi(X_{t}) \, dt +  \nabla\cdot \cM(X_{t})\,dt + \sqrt{2 \cM(X_{t})}\, dW_{t},
\end{equation}
where $\Psi(x)$ denotes the free energy and $\mathcal{M}(x)$ the effective diffusion tensor:  \begin{equation}\label{e:free-energy}
\Psi(x) = - \ln Z(x), \quad  Z(x) = \int_{\mathbb{T}^d}\cdots\int_{\mathbb{T}^d}e^{- V(x, y_1,\ldots, y_N)/\sigma}\,dy_N\ldots dy_1,
\end{equation}
and
\begin{equation}\label{e:Meff}
\mathcal{M}(x) = \frac{\sigma}{Z(x)}\int_{\mathbb{T}^d}\cdots\int_{\mathbb{T}^d}(I + \nabla_{y_N}\theta_N)\cdots(I+\nabla_{y_1}\theta_1)e^{- V(x,y_1,\ldots,y_N)/\sigma}\,dy_N\cdots dy_1.
\end{equation}
The corrector fields $\big\{\theta_1(x), \dots \theta_N(x) \big\}$ are
defined recursively as follows: let $\theta_{N-k}$ be the solution of
\begin{equation}\label{e:corrector-1}
	\nabla_{x_{N-k}}\cdot(\mathcal{M}_{N-k}(x, y_0, \ldots, y_{N-k})(\nabla_{x_{N-k}}\theta_{x_{N-k}}(x,y_1,\ldots, y_{N-k})+I)) =0, \quad y_{N-k} \in \mathbb{T}^d,
\end{equation}
where for $1 \leq k < N$:
\begin{equation}\label{e:corrector-2}
\mathcal{M}_{N-k}(x, y_1,\ldots, y_{N-k}) = \int_{\mathbb{T}^d}\cdots\int_{\mathbb{T}^d}(I + \nabla_N\theta_N)\cdots(I+\nabla_{N-k+1}\theta_{N-k+1})e^{- V/\sigma}\,dy_N\ldots dx_{N-k+1},
\end{equation}
and  $\mathcal{M}_N(x, y_1,\ldots, y_N) = e^{-\beta V(x, y_1,\ldots, y_N)}I$. It is possible to show that the effective diffusion tensor is positive definite, uniformly in $x \in \R^d$ and to obtain upper and lower bounds on $M$.

The homogenized dynamics $X(t)$ is exponentially ergodic and reversible with respect to the invariant distribution
\begin{equation}\label{e:inv-distr}
\rho(x) = \frac{Z(x)}{\overline{Z}}, \quad \overline{Z} = \int_{\R^d} Z(x) \, dx.
\end{equation}
The generator of the homogenized dynamics, which is a self-adjoint operator in $L^2(\R^d, \rho(x))$ can be written in the form
\begin{equation}\label{e:generator-homog}
\mathcal{L} \cdot = \frac{1}{Z(x)}\nabla_x\cdot\left(Z(x)\mathcal{M}(x)\nabla_x \cdot \right).
\end{equation}
We remark that $\overline{Z}$ in~\eqref{e:inv-distr} is the partition
function of the full dynamics $X_{t}^{\eps}, \, Y^{\eps}_{t}$ and requires
the calculation of an integral over $\R^{d} \times \T^{d}$. It can be shown that  invariant distribution of the homogenized dynamics is the weak limit of the invariant distribution \eqref{e:xeps_inv_meas_ddim} of $X(t)^{\eps}$. This follows
from properties of periodic functions~\cite[Ch. 2]{cioran}.

The coarse-grained equation~\eqref{e:homog} that we derive here provides us
with a rigorous derivation of the free energy~\eqref{e:free-energy} for
systems with strong scale separation which can be used, in turn, to compute
equilibrium coarse-grained quantities~\cite{BlancLeBrisLegollPatz2010}. On
the other hand, the homogenized dynamics~\eqref{e:homog} is the most general
form of a reversible diffusion process with respect to the invariant
distribution~\eqref{e:inv-distr}, see~\cite[Sec 4.6]{Pavl2014} and can be
used to study time-dependent phenomena such as bifurcations and noise-induced
transitions.  Indeed, an important point to note is that for the case of
nonseparable potentials, as given by  \eqref{e:pote-form}, all scales are
fully coupled in the hierarchy of Poisson
equations~\eqref{e:corrector-1}-\eqref{e:corrector-2}. As a result of this,
even through the noise in the original dynamics~\eqref{e:gen-sde} is due to
thermal fluctuations and is hence additive, the noise in the coarse-grained
dynamics is multiplicative, something which, as it is well known and as was
emphasized in the Introduction, can lead to noise-induced
transitions~\cite{MackeyLongtinLasota1990}. These points will be elucidated
in Sections \ref{sec:pitchfork} and \ref{subsec:stabiliz}.

A final remark is that the noise in the coarse-grained dynamics becomes
additive, when the potential~\eqref{e:pote-form} is separable, i.e.
\begin{equation}\label{e:pot-additive}
V^{\eps}(x) = \sum_{n=0}^N V_n \left(\frac{x}{\eps^n} \right),
\end{equation}
(a potential that could be achieved by design in a physical setting, and
hence also realistic), a surprising result and perhaps counterintuitive as
one might expect that coarse graining always leads to multiplicative noise in
the effective description. Now the Poisson
equations~\eqref{e:corrector-1}-\eqref{e:corrector-2} can be solved in a
hierarchical fashion, and the homogenized equation is of the
form~\eqref{e:homog}, but with a constant effective diffusion tensor. For
illustrative purposes we present the formulas for $N=1$~\cite{PavlSt06}, in
which case the effective dynamics is given by the SDE:
\begin{equation}
 d X_t = - \cM \nabla V_0(X_t) \, dt + \sqrt{2 \sigma \cM} \, d W_t,
\label{e:lim_sde}
\end{equation}
where
\begin{equation}
\cM = \int_{\T^d} \left( I + \nabla_y \theta(y) \right)  \left( I + \nabla_y \theta(y)
\right)^T \, \mu(dy) \label{e:coeffs}
\end{equation}
and
\begin{equation}
\mu(dy) = \rho(y) dy = \frac{1}{Z} e^{- \sigma^{-1} V_1(y)} \, dy, \quad Z = \int_{\T^d}
e^{- \sigma^{-1} V_1(y)} \, dy. \label{e:gibbs_torus}
\end{equation}
The field $\phi(y)$ is the solution of the Poisson equation
\begin{equation}
- \LL_0 \theta(y) = -\nabla_y V_1(y), \quad \LL_0 := - \nabla_y V_1(y) \cdot \nabla_y + \sigma
\Delta_y, \label{e:cell}
\end{equation}
with periodic boundary conditions.
%
%
\subsection{The homogenized equation in 1D}
It is well known that homogenized coefficients in 1D can be computed
explicitly, up to quadratures~\cite[Sec. 12.6.1, Sec. 13.6.1]{PavlSt08}. This
is the case for the $N-$scale homogenization problem that we consider in this
study. In \ref{app:1d-deff} we show how, by solving the family of Poisson
equations in~\eqref{e:corrector-1}-\eqref{e:corrector-2} and by using
formula~\eqref{e:Meff} we obtain the following formula for the effective
diffusion coefficient
\begin{equation}\label{e:homog-1d}
\cM(x) = \frac{1}{Z_+(x) Z_-(x)}, \quad Z_{\pm}(x) =  \int_{\mathbb{T}^N}\mathrm{e}^{\mp V(x, y_1,\ldots, y_N)/\sigma}dy_1\dots dy_N,
\end{equation}
see also \cite{Zwan88}.  Of course, the explicit calculation of the effective diffusion coefficient in one dimension using~\eqref{e:homog-1d} requires the calculation of the partition functions $Z_{\pm}(x)$. In Section~\ref{sec:piecewise} we show how these multiple integrals can be calculated analytically for the case of an $N-$scale potential that is piecewise linear at all scales.

In the following sections we will consider different examples of multiscale potentials in one dimension to study the interplay between noise and the  multiscale structure of the potential. Our goal, in particular, is to understand how the microscopic fluctuations can affect the global dynamics of the system.
%
%
\section{A pitchfork bifurcation: Noise-induced hysteresis}
\label{sec:pitchfork}

Consider the following SDE:
\begin{equation}
\label{eq:pitchfork}
  d X(t) = (\alpha X(t) - X(t)^3)dt + \sqrt{2\sigma}dW(t),
\end{equation}
the deterministic part of which is the normal form for a supercritical pitchfork bifurcation.  For $\alpha < 0$ there is a single stable equilibrium at $0$ while for $\alpha > 0$ there is an unstable equilibrium at $x = 0$ and two stable equilibria at $x=\pm\sqrt{\alpha}$.  The system described by  (\ref{eq:pitchfork}) has a potential $V_0(x ; \alpha) = -\frac{\alpha}{2}x^2 + \frac{1}{4}x^4$.

We define a two-scale potential by introducing a rapid fluctuation on the bifurcation parameter $\alpha$ so the potential reads:
\begin{equation}
V(x, y; \alpha) = \frac{1}{4}x^4 -\left[\frac{\alpha + g_\delta(x)\sin\left(2\pi y\right)}{2}\right] x^2,
\end{equation}
which in turn can be rewritten as:
\begin{equation}
\label{eq:pitchfork_potential}
V(x, y; \alpha) = V_0(x;\alpha) -\frac12g_\delta(x) x^2\sin\left(2\pi y\right),
\end{equation}
where we have introduced a decaying function with the properties $g_\delta(0) =1$ and
$g_\delta(x\gg \delta)\to 0$. Note that the main purpose of this function is to ensure
that the microscopic fluctuations are confined within the region in which the macroscopic
potential varies.  An illustation of the above potential for the case of $\alpha = 0.5$ is shown in  figure \ref{Fig: potentials}.

We start by studying the equilibrium properties of  (\ref{eq:pitchfork}). The ergodic distribution of the homogenised dynamics is given by  (\ref{e:inv-distr}), which after including the potential given by  (\ref{eq:pitchfork_potential}), we obtain:
\begin{equation}
\label{Eq: rho 0 2s}
\rho(x) = \frac{1}{Z}e^{- V_0(x;\alpha)/\sigma} I_0\left(\frac{g_\delta  x^2}{2\sigma}\right),
\end{equation}
where $I_0(\cdot)$ is the modified Bessel function of the first kind which depends on both the position $x$ and the noise intensity $\sigma$, and it is a correcting term coming from the microscopic fluctuations---note that if $g_\delta =0$ we recover the Gibbs measure of the unpertubed macroscopic system. Therefore, we can see that the microscopic fluctuations are able to modify the equilibrium points of the system (i.e.~the maxima of $\rho(x)$) and these are controlled by the noise intensity.

\begin{figure}[t]
\centering
\includegraphics[width=1\textwidth]{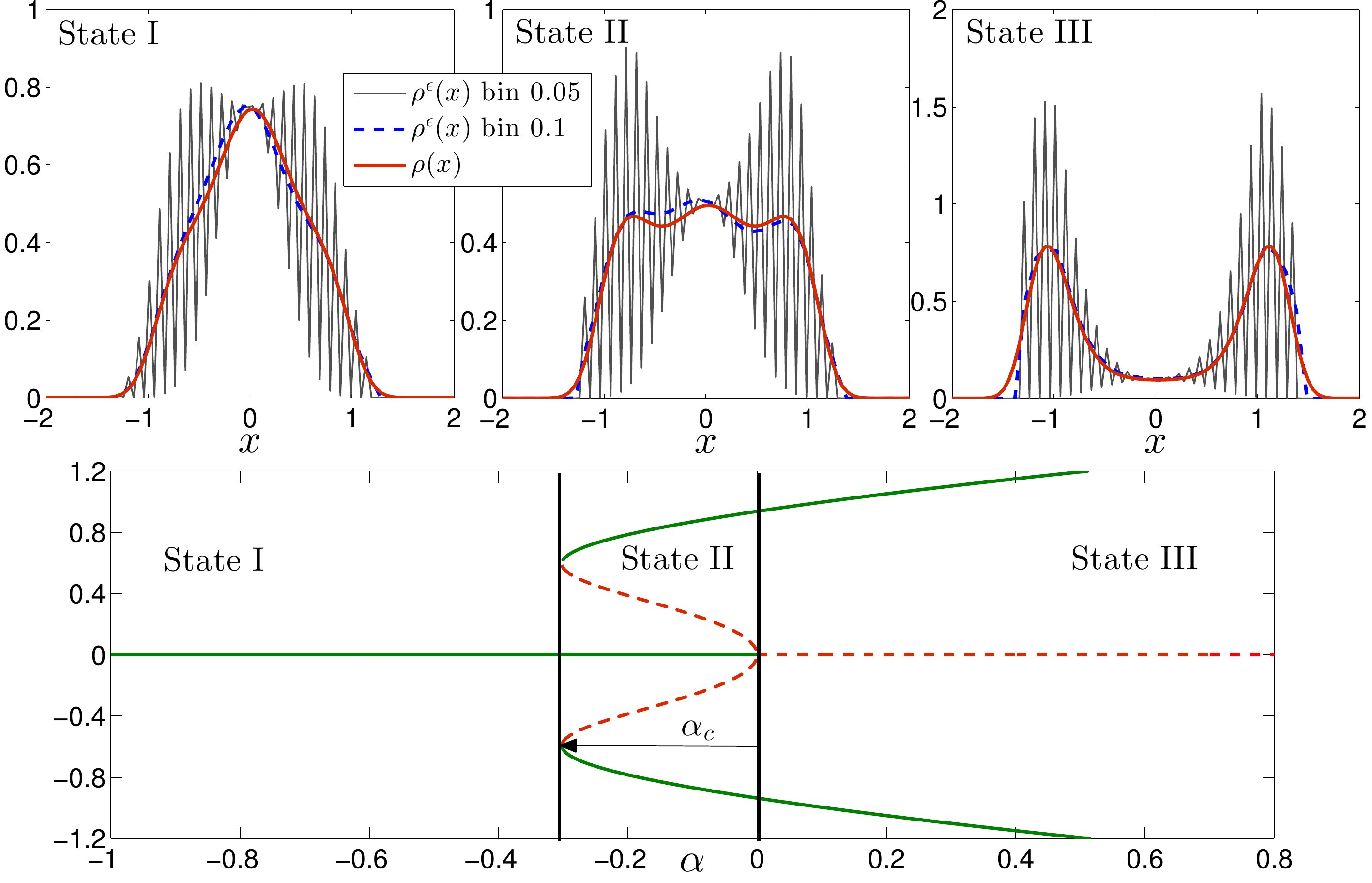}
\caption{Top panels show histograms generated over a long sample of the SDE (\ref{eq:pitchfork}) with the two-scale
potential (\ref{eq:pitchfork_potential}) with  $\sigma = 0.1$, $\epsilon =
10^{-1}$, and using bins of size $0.05$ (thin gray line) and $0.1$ (dashed
blue line), for the three different dynamical states: State $\mathrm{I}$ with
$\alpha = -0.5<\alpha_c$, state $\mathrm{II}$ $\alpha = -0.25\in[\alpha_c,0]
$, and state $\mathrm{III}$ with $\alpha= 0.25>0$. In all three panels, the
thick red line shows analytical homogenised solution given by (\ref{Eq: rho 0
2s}). The bottom panel shows the corresponding bifurcation diagram for
reference with the three states demarcated with vertical solid
lines.}\label{fig:SDE 2s}
\end{figure}

To quantify this effect, we construct the bifurcation diagram of the solution for different values of $\sigma$.  To this end we look at the equilibria $x_s$ of the above function $\rho(x)$ which are given by the solution of the following equation:
\begin{equation}
\label{Eq: equ 2s}
-x_s^3 +x_s\left[ \alpha +\frac{I_0'(x_s^2g_\delta/2\sigma)}{I_0(x_s^2g_\delta/2\sigma)}\right]=0.
\end{equation}
\begin{figure}[t]
\centering
\includegraphics[width=1.0\textwidth]{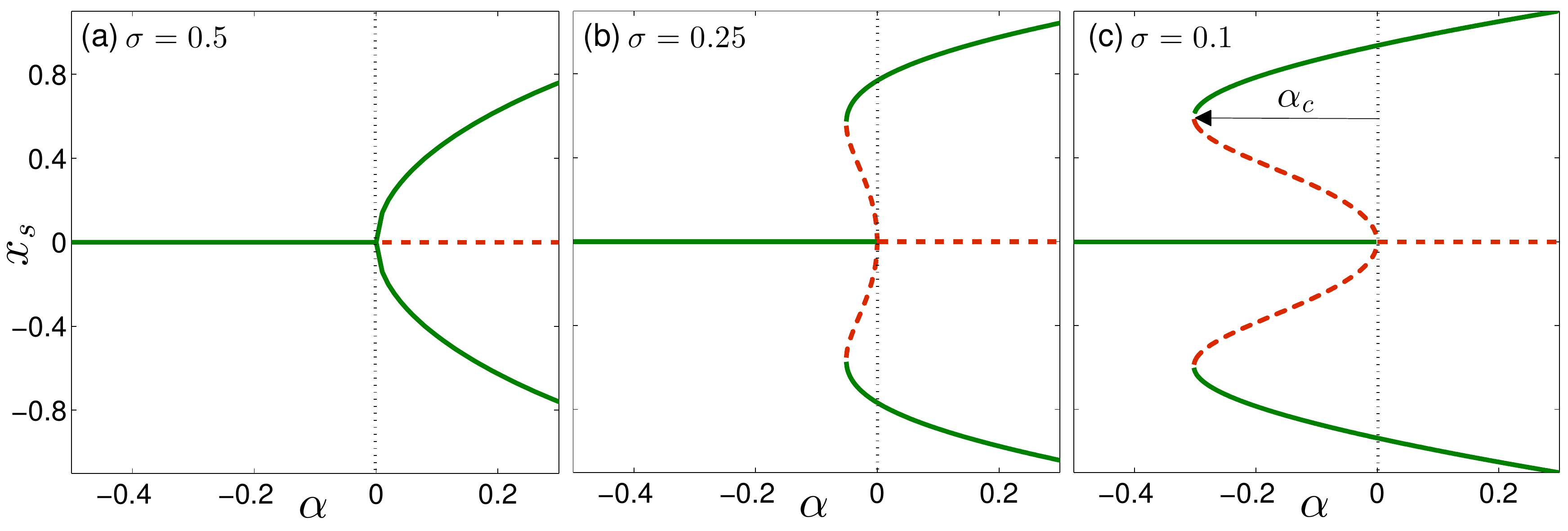}
 \caption{Bifurcation diagram of the two-scale potential given by  (\ref{eq:pitchfork_potential}) obtained by solving   (\ref{Eq: equ 2s}) for (a) $\sigma = 0.5$, (b) $\sigma = 0.2$, and (c) $\sigma = 0.1$.}
\label{fig:BD 2s}
\end{figure}
The results are presented in  figure \ref{fig:BD 2s} where we can see that
for sufficiently large values of the noise level, the long-time behaviour of
the macroscopic system demonstrates the same qualitative behaviour as the
unperturbed case with  a \emph{supercritical} pitchfork bifurcation occurring
when $\alpha=0$ [cf.~ Figure \ref{fig:BD 2s}(a)]. However,  as the  noise
intensity decreases,  the behaviour becomes qualitatively different. Indeed,
for some critical value of $\sigma$ the pitchfork bifurcation becomes
\emph{subcritical} and two saddle-node bifurcations symmetric about the
$x$-axis arise along the negative axis,  giving rise to three stable and two
unstable branches.  As $\alpha$ passes $0$ the central stable branch becomes
unstable [cf.~Figs.~\ref{fig:BD 2s}(b,c)]. In this scenario, we can identify
three different dynamic states: (I) for $\alpha<\alpha_c<0$ zero is a stable
solution of the system, (II) for $\alpha_c<\alpha<0$, in addition to zero,
there are two other non-zero stable solutions, and (III) for $\alpha>0$, zero
becomes unstable and there are two non-zero stable solutions. We note that
this system gives rise to a \emph{hysteresis} loop and the macroscopic system
will not follow the same equilibrium branch for $\alpha$ increasing as when
$\alpha$ is decreased.
\\\\
To further illustrate the transitions that arise in our multiscale system, we
simulate the evolution of a Brownian motion in the two-scale potential
(\ref{eq:pitchfork_potential}).  We choose $\sigma = 0.1$, $\epsilon = 0.1$
and $g_{\delta}$ to be a smooth mollification of the indicator function over
$[-10,10]$.   We approximate the SDE numerically using a standard
Euler-Maruyama discretisation with step size $\Delta t = 10^{-4}$.  In Figure
\ref{fig:SDE 2s} we plot histograms generated from $10$ independent runs each
of $10^9$ timesteps, for $\alpha=-0.5$, $-0.25$ and $0.25$, respectively. The
choices of $\alpha$ correspond to the dynamics before the bifurcation point,
close to bifurcation point and after bifurcation, respectively, as
illustrated in the bottom panel of figure \ref{fig:SDE 2s}. They correspond
to the three dynamical states defined above. In each case, the red line
denotes the exact stationary distribution $\rho(x)$ given by (\ref{Eq: rho 0
2s}). The thin gray line denotes a normalized histogram, generated from the
samples lying in $[-2,2]$ with the size of each bin taken to be $0.05$.  We
see that the approximated density exhibits large fluctuations around
$\rho(x)$.  This is to be expected since the stationary density
$\rho^\epsilon(x)$ does not converge pointwise to $\rho(x)$, but only in the weak sense.  Indeed, increasing the size of the  histogram bins $0.1$, as depicted by the dashed blue line, we see much better agreement in each case.

%
\subsection{Extension to $N$ scales}
\begin{figure}[t]
\centering
\includegraphics[width=1.0\textwidth]{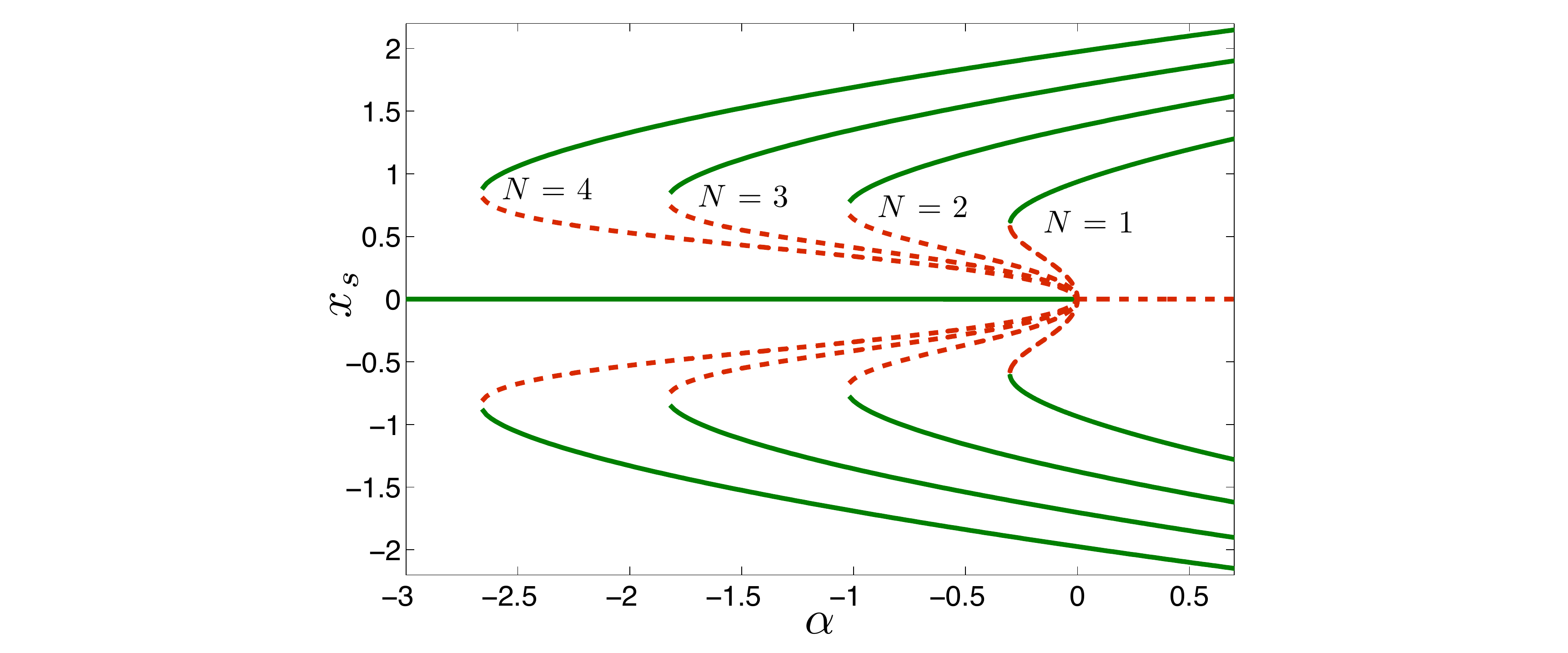}\\
\caption{Bifurcation diagram for different values of the number of microscopic scales $N=1,2,3,4$ (from left to right) in the potential defined in  (\ref{eq:pitchfork_potential N}).}
\label{Fig:Bif N scales}
\end{figure}

A natural extension in the two-scale potential of  (\ref{eq:pitchfork_potential}) is to add more microscopic scales, say up to $N$, so that the new potential is of the form:
\begin{equation}
\label{eq:pitchfork_potential N}
V(x, y; \alpha) = V_0(x;\alpha) -\frac12g_\delta(x)  x^2 \sum_{n=1}^{N}\sin\left(2\pi y_n\right).
\end{equation}
for $y_n = x/\epsilon^n$. In this case, the stationary distribution reads:
\begin{equation}
\label{Eq: rho 0 Ns}
\rho(x) = \frac{1}{Z}e^{-\beta V_0}\left[I_0\left(\frac{x^2g_\delta}{2\sigma}\right)\right]^N,
\end{equation}
and  (\ref{Eq: equ 2s}) becomes:
\begin{equation}
\label{Eq: equ Ns}
-x_s^3 +x_s\left[ \alpha +N\frac{I_0'(x_s^2g_\delta/2\sigma)}{I_0(x_s^2g_\delta/2\sigma)}\right]=0.
\end{equation}
By computing again the equilibrium points and constructing the corresponding
bifurcation diagram, we observe that the transition from supercritical to
subcritical is in fact enhanced with the number of scales $N$ (see figure
\ref{Fig:Bif N scales} for the case with $\sigma=0.1$).

%
\begin{figure}[t]
\centering
\includegraphics[width=1.0\textwidth]{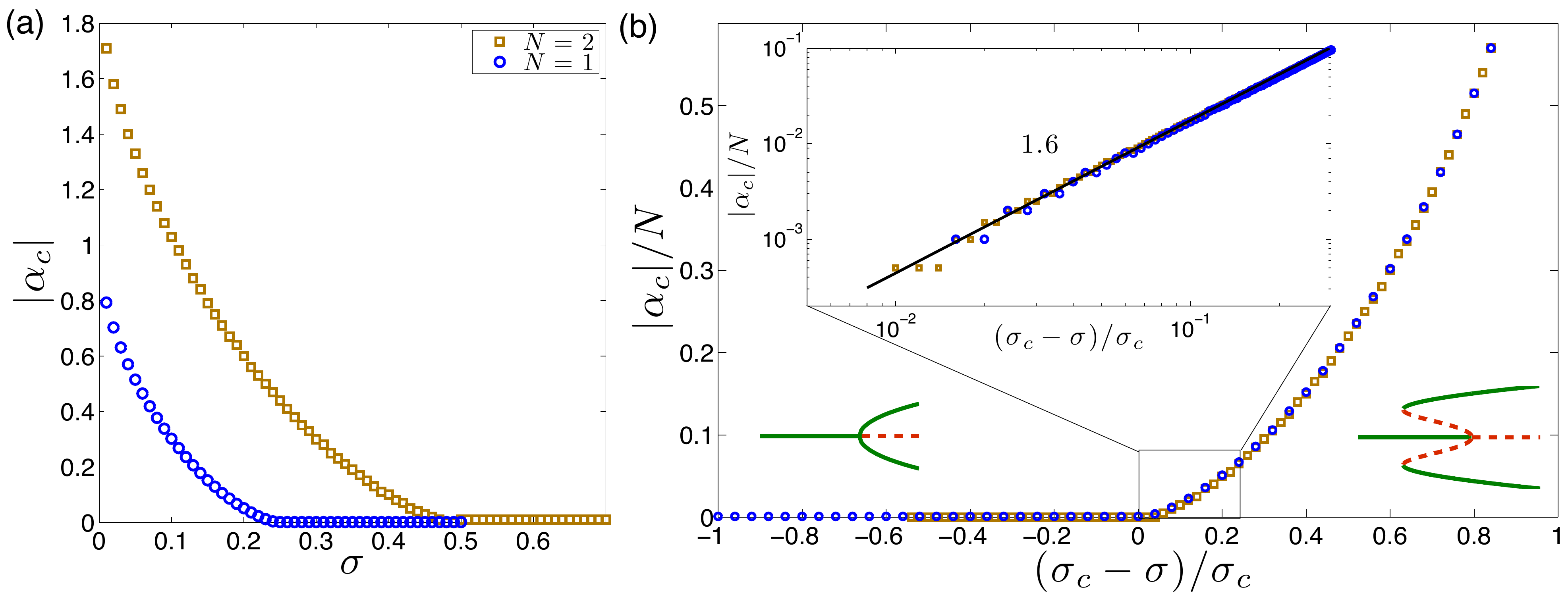}\\
\caption{Results obtained by using the multiscale potential defined in  (\ref{eq:pitchfork_potential N}).
(a) Transition from supercritical ($\vert\alpha_c\vert =0$) to subcritical
($\vert\alpha_c\vert >0$) as the noise intensity is decreased. (b) Rescaled
$\sigma$ and $\vert\alpha_c\vert$ show data collapse into a single curve. The
inlet shows a log-log plot where we can identify a power-law behaviour close
to criticality with exponent $\gamma = 1.6$. The corresponding
bifurcation diagrams for the supercritical and subcritical cases are also
shown for reference.} \label{Fig: Transition 2s}
\end{figure}
\begin{figure}
\centering
\includegraphics[width=1.0\textwidth]{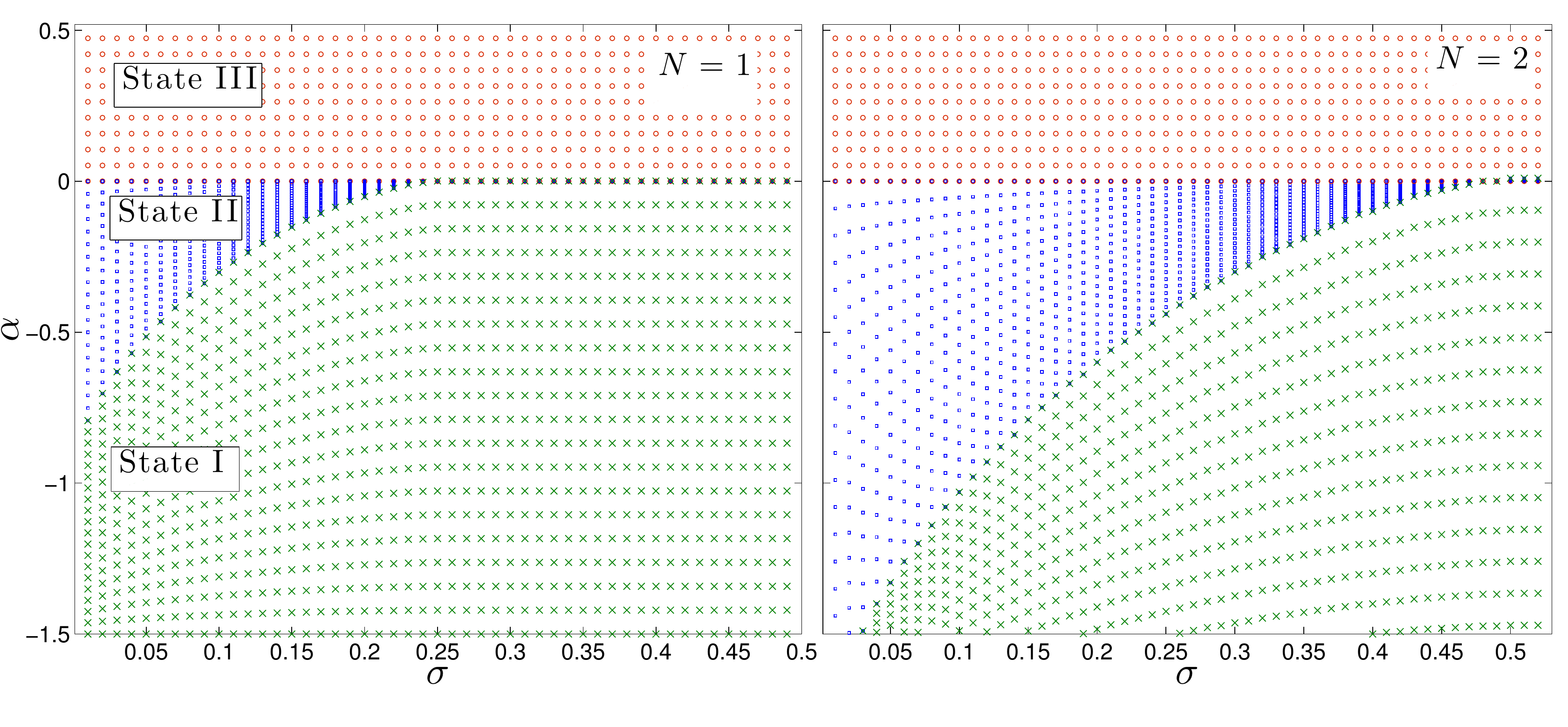}\\
\caption{Phase diagram showing the three dynamical states $\mathrm{I}$ (green) , $\mathrm{II}$ (blue), and $\mathrm{III}$ (red), which are observed in the potential given
in  (\ref{eq:pitchfork_potential N}) and defined in  figure \ref{fig:BD 2s}
for two values of the number of microscopic states $N$.} \label{Fig:phase
diagram}
\end{figure}
To quantify the transition from a supercritical to a subcritical pitchfork
bifurcation observed when the noise intensity is decreased (cf.~ figure
\ref{fig:BD 2s}), and how this depends on the number of microscopic scales
$N$, we take the absolute value of $\alpha_c$ defined in the bottom panel of
figure \ref{fig:SDE 2s} to play the role of  an order parameter of the
transition, such that the bifurcation is supercritical for
$\vert\alpha_c\vert = 0$ and subcritical for $\vert \alpha_c\vert>0$. The
results are depicted in  figure \ref{Fig: Transition 2s}(a), where we can
observe that $\vert\alpha_c\vert$ becomes zero at some critical value
$\sigma_c$ which depends on the number of microscopic scales. Indeed,  by
considering values of $x_s\ll 1$ in  (\ref{Eq: equ Ns}), we can expand the
Bessel functions ($I'_0(x)/I_0(x)\sim x/2$ for $x\ll 1$) and we can see that
the two non-zero solutions exist for values of $\sigma$ which are below the
critical value:
\begin{equation}
\sigma_c = \frac{N}{4},
\end{equation}
which defines the critical point.  By taking now the rescaled variables:
\begin{equation}
X \equiv \frac{\sigma_c-\sigma}{\sigma_c}, \qquad Y \equiv\frac{\vert\alpha_c\vert}{N},
\end{equation}
we can observe that all data collapse into a single curve, and close to critical point the transition is characterized by a power-law behaviour:
\begin{equation}
Y\sim \vert X\vert ^\gamma,
\end{equation}
with $\gamma=1.6$ [cf. figure~\ref{Fig: Transition 2s}(b)]. In addition, we
present in  figure \ref{Fig:phase diagram} a phase diagram on the plane
($\sigma,\alpha$) where we can see how the different dynamical states (I),
(II), and (III) defined above depend on the number of microscopic scales $N$.

\section{Noise-induced stabilization}
\label{subsec:stabiliz}
%
%
%

\begin{figure}
\centering
\includegraphics[width=1.0\textwidth]{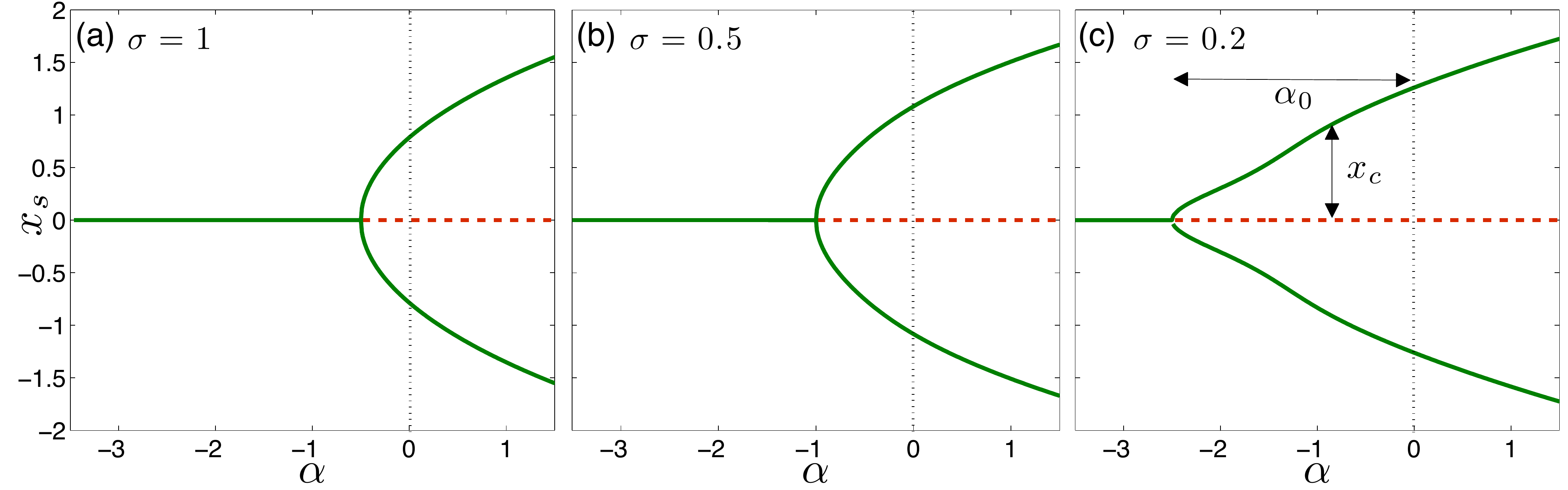}\\
\caption{Bifuraction diagram for the three-scale potential and
for three different values of the noise strength, namely $\sigma=0.2$ (a),
$\sigma =0.5$ (b), and $\sigma =1$ (c).}
\label{Fig:bif 3s}
\end{figure}

We consider now a tilted three-scale quartic potential:
\begin{equation}
V(x,y_1,y_2;\alpha) = \frac14x^4 - \bigg[\frac{\alpha+g_\delta\sin{(2\pi y_1)}}{2}\bigg] x^2 + g_\delta\lambda\sin{(2\pi y_2)}x.
\end{equation}
which we rewrite as:
\begin{equation}
\label{Eq:3s pot}
V(x,y_1,y_2;\alpha) = V_0(x;\alpha) - \frac12g_\delta x^2\sin{(2\pi y_1)} + g_\delta\lambda x\sin{(2\pi y_2)}.
\end{equation}
An example of this potential is shown in  figure \ref{Fig: potentials}. The parameter $\lambda$ above has been introduced to connect it with the potential presented in the previous section which is recovered when $\lambda=0$. As before, we start by looking at the stationary distribution which we find to be:
\begin{equation}
\label{Eq: rho 0 3s}
\rho(x) = \frac{1}{Z}e^{-\beta V_0}I_0\left(\frac{x^2g_\delta}{2\sigma}\right)I_0\left(\frac{x g_\delta\lambda}{\sigma}\right).
\end{equation}
Figure \ref{Fig:3s Num Theo} shows numerical computations of both the stationary distribution obtained by solving the multiscale SDE  with the potential given in  (\ref{Eq:3s pot}) and the above analytical solution $\rho(x)$ for different values of $\sigma$ and $\alpha$, observing an excellent agreement in all cases. In addition, we look at the equilibria of $\rho(x)$ which are given by the following equation:
\begin{figure}[t]
\centering
\includegraphics[width=1.0\textwidth]{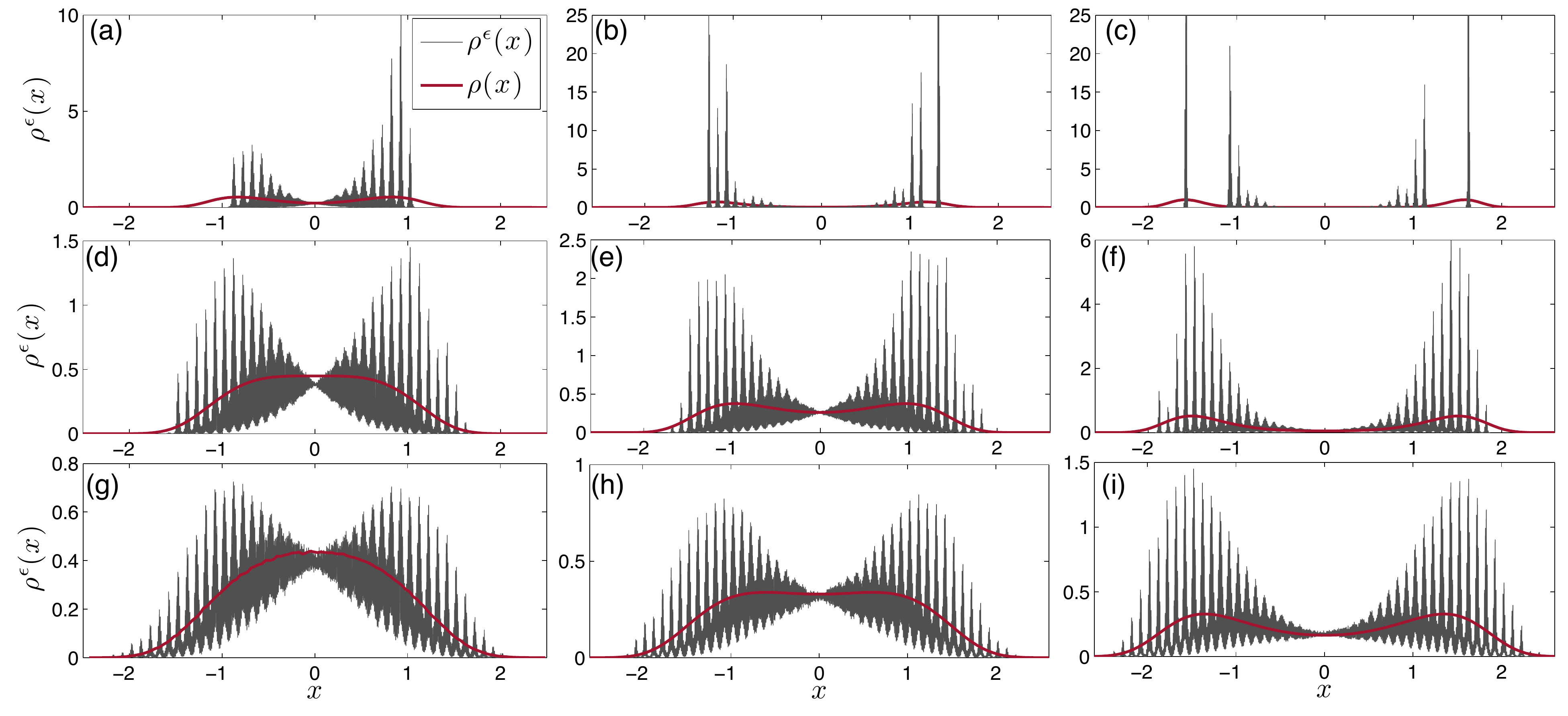}
\caption{Numerical solution of the stationary distribution $\rho^{\epsilon}(x)$ by solving the multiscale SDE  with the potential given in (\ref{Eq:3s pot}) and the analytical solution $\rho(x)$ given by  (\ref{Eq: rho 0 3s}) (solid red line) for different values of the noise strength and the parameter $\alpha$. (a), (b), and (c)
correspond to $\sigma=0.2$ for $\alpha=-1$, $-0.2$ and $1$, respectively. (d), (e), and (f) correspond to $\sigma=0.5$ for $\alpha=-1$, $-0.2$ and $1$, respectively; and (g), (h), and (i) correspond to $\sigma=1$ for $\alpha=-1$, $-0.2$ and $1$, respectively.}
\label{Fig:3s Num Theo}
\end{figure}
\begin{equation}\label{Eq:critical points}
-x_s^3 +x_s \left[ \alpha +\frac{I_0'(x_s^2g_\delta/2\sigma)}{I_0(x_s^2g_\delta/2\sigma)}\right]+
\lambda\frac{I_0'(x_s g_\delta\lambda/\sigma)}{I_0(x_s g_\delta\lambda/\sigma)}=0,
\end{equation}
from which we construct the bifurcation diagram, shown in  figure \ref{Fig:bif 3s} for different values of the noise level $\sigma$. It is interesting to note that for this potential the transition to subcritical bifurcation is not observed but rather the supercritical pitchfork bifurcation is being shifted to the left as the noise intensity is decreased. Moreover, it is remarkable that the macroscopic unperturbed behaviour is only recovered for sufficiently large values of $\sigma$.
To make this statement more precise, we look at the value of $\alpha_0$  which is defined as the value where the pitchfork bifurcation occurs (see  figure \ref{Fig:bif 3s}) and which satisfies the following condition:
\begin{equation}
\partial^2_x \rho(x;\alpha_0)\vert_{x=0}=0,
\end{equation}
giving rise to:
\begin{equation}
\alpha_0 = -\frac{\lambda}{2\sigma}.
\end{equation}
We hence conclude that for the three-scale potential, the case $\alpha_0=0$,
which corrsponds to the standard supercritical bifurcation, is only achieved
when $\sigma\to \infty$ [we note also that for the unperturbed macroscopic
dynamics ($g_\delta = 0$) and for the case of a two-scale potential analyzed
in the previous section, $\lambda=0$, we have $\alpha_0=0$ independently of
$\sigma$].

\begin{figure}
\centering
\includegraphics[width=1.0\textwidth]{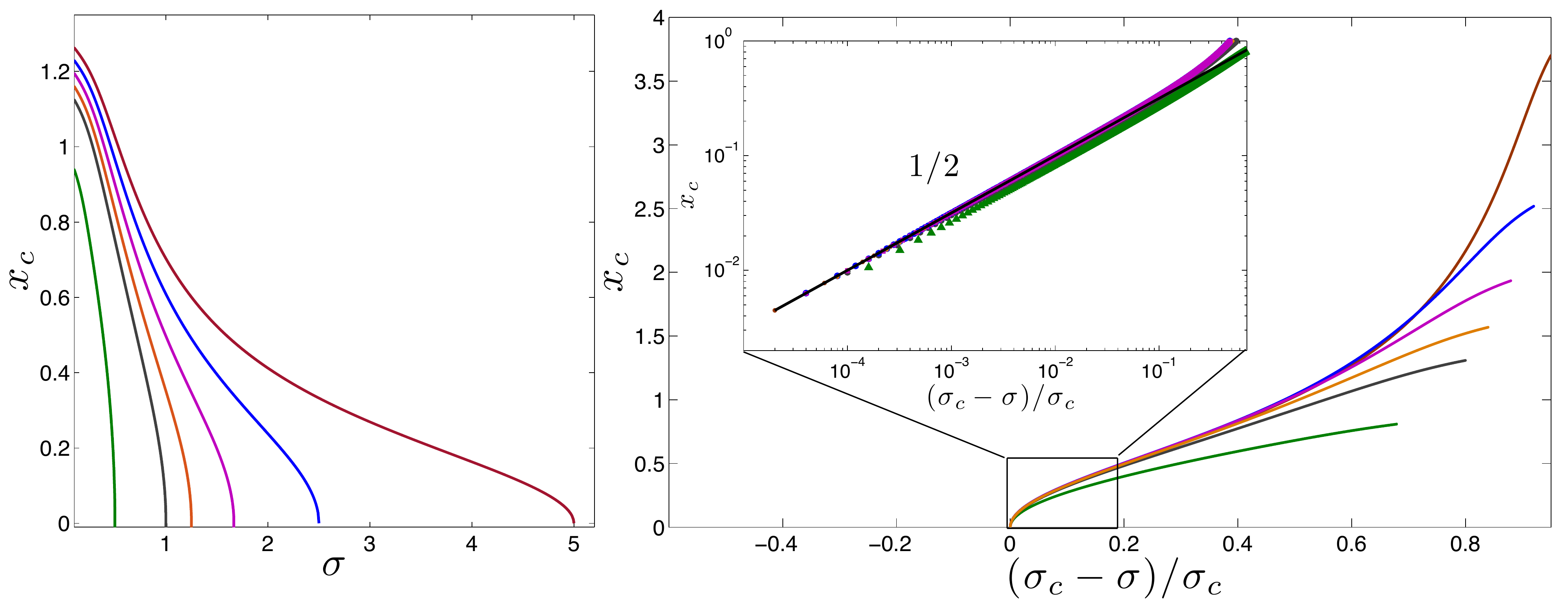}\\
\caption{Position $x_c$ of the local maximum of the PDF as function
of the noise strength $\sigma$ for different values of the parameter
$\alpha$, namely $\alpha =-1$, $-0.5$, $-0.4$, $-0.3$, $-0.2$, $-0.1$ (curves
from left to right). Panel on the right show the position of the maximum with
respect to the the rescaled variable $(\sigma_c-\sigma)/\sigma_c$. When the solution
is plot in a log-log scale (inlet panel) we can observe a power-law behaviour with exponent
$1/2$.}
\label{Fig:max PDF}
\end{figure}

An important consequence of the fact that $\alpha_0$ depends on $\sigma$ is
that the stability of the zero solution can be tuned by changing the noise
strength. Indeed, if we take a fixed (negative) value $\alpha$, the zero
solution will be unstable for values of $\sigma$ which are below the critical
value:
\begin{equation}
\sigma_c = \frac{\lambda}{2\vert \alpha\vert},
\end{equation}
and stable otherwise. How this transition is approached as we increase the value of $\sigma$ can be studied by looking at the position
of the local maximum, say $x_c$, of the stationary distribution $\rho(x)$ which is a solution of  (\ref{Eq:critical points}) [see figure \ref{Fig:bif 3s}(c)
for the definition of $x_c$ for a given value of $\alpha$]. For a fixed value of $\alpha$ we then have that the zero solution is stable when $x_c= 0$ and unstable when $x_c>0$. We can therefore define $x_c$ to be the order parameter of this transition. 
Figure \ref{Fig:max PDF} shows how the position of one of the two maxima of
the PDF approaches the value of zero as $\sigma$ is increased and for
different values of the chosen $\alpha$, where we can see that near the
critical point the solution exhibits a power-law behaviour of the form:
\begin{equation}
x_c\sim \vert X\vert^\gamma,
\end{equation}
with $\gamma=1/2$. Indeed, we can verify this behaviour analytically if we
look at the solutions given by (\ref{Eq:critical points}) in the limit of
$x_c\ll 1$ for which we can expand the Bessel functions [$I'_0(x)/I_0(x)\sim
x/2$ for $x\ll 1$]. Expanding around the critical point $\sigma = \sigma_c -
\delta \sigma$ yields to leading order in $\delta\sigma$ that $x_c\sim
\vert\delta\sigma/\sigma_c\vert^{1/2}$.

%

\section{Brownian motion in a piecewise linear self-similar potential}
\label{sec:piecewise}

\begin{figure}[!t]
  \centering
 \includegraphics[width=1.0\textwidth]{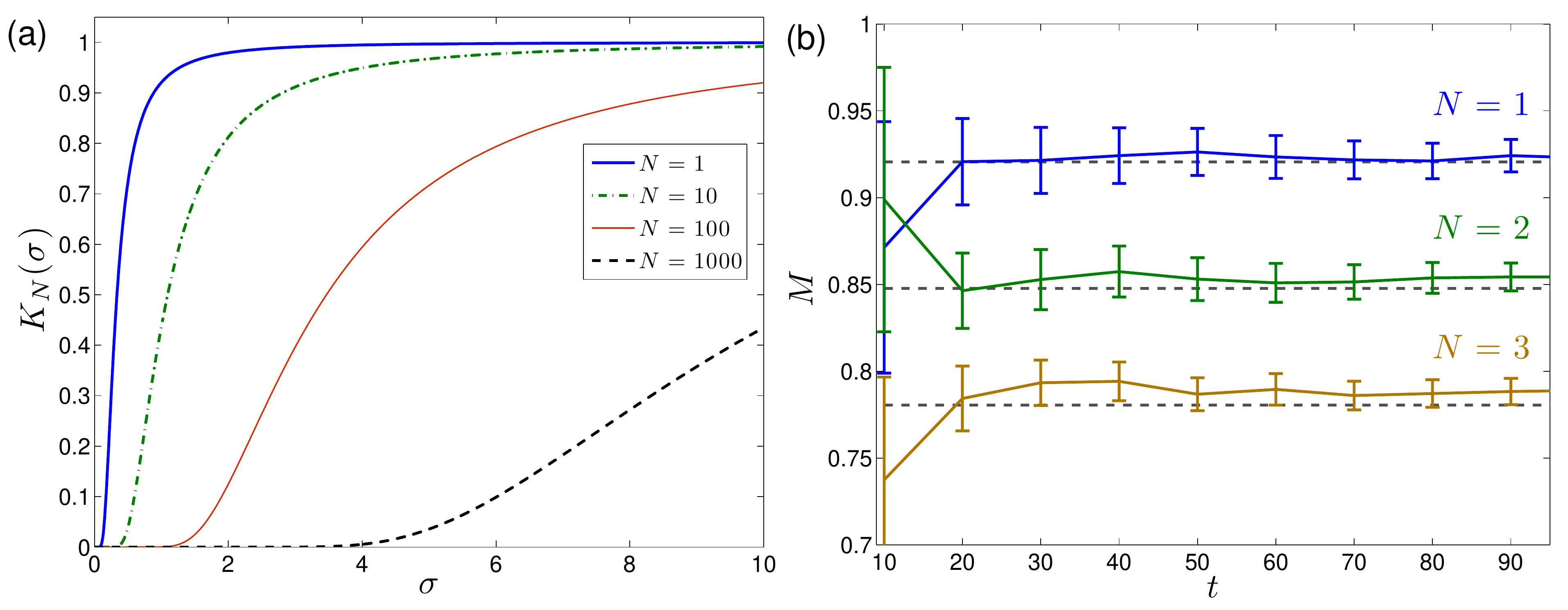}
  \caption{(a) Plot of the retardation factor $K_N(\sigma)$ over $\sigma$ for different values of $N$. (b) Effective diffusion coefficient approximated from numerical simulations for $N=1$, $2$ and $3$ compared to the homogenized diffusion coefficient $M(\sigma)$ predicted by (\ref{e:pw M}) (dashed lines).}
  \label{fig:potential_pw}
\end{figure}

In this section, we consider  a piecewise-linear $N$-scale separable potential given by
\begin{equation}
\label{eq:pw pot}
  V^\epsilon(x) = V_N\left(\frac{x}{\epsilon}, \ldots, \frac{x}{\epsilon^N}\right) = S\left(\frac{x}{\epsilon}\right) +  S\left(\frac{x}{\epsilon^2}\right) + \ldots +  S\left(\frac{x}{\epsilon^N}\right),
\end{equation}
where
\begin{equation}
 S(x) = \left\{ \begin{array}{ll}
        2x & \mbox{ if } x \mbox{ mod } 1 \in [0, \frac{1}{2});\\
        2 - 2x & \mbox{ if } x \mbox{ mod } 1 \in [\frac{1}{2}, 1),\end{array} \right.
\end{equation}
for fixed $N \in \mathbb{N}$ and $\epsilon > 0$.  Since we are dealing with a
separable potential with no large-scale component, we know from the results
presented in Section~\ref{sec:homog} that the coarse-grained dynamics is
purely diffusive, i.e. the coarse-grained Fokker-Planck equation is the heat
equation:
 \begin{equation}
\frac{\partial F_0(x, t)}{\partial t}\, = M(\sigma) \frac{\partial^2 F_0(x, t)}{\partial x^2},
\end{equation}
where $M$ is a constant effective diffusion tensor. Since the fast scale
fluctuations in the potential are separated, as it is described in
\ref{app:1d-deff} we can easily obtain the constant effective diffusion from
(\ref{eq:eff_diff_1d_separable}):
 \begin{align}
 \label{e:pw M}
 M(\sigma) &=  \sigma\left(\int_{\mathbb{T}} e^{- S(z)/\sigma}\, dz \cdot \int_{\mathbb{T}} e^{ S(z)/\sigma}\, dz\right)^{-N} \nonumber\\
  & =  \sigma\left[\sigma^{2}\left(1 -e^{-1/\sigma }\right)\cdot \left(e^{1/\sigma }-1\right)\right]^{-N} \nonumber\\
  & =  \frac{\sigma}{\left[2\sigma^2\left(\cosh \left(\frac{1}{\sigma}\right)-1\right)\right]^{N}}.
 \end{align}
%
The $N$-scale perturbations have a retarding effect on the motion which is
amplified as $N$ increases, a consequence of the increased complexity of the
potential. This is captured in the scalar term
$K_N(\sigma)=\left[2\sigma^2\left(\cosh
\left(\frac{1}{\sigma}\right)-1\right)\right]^{-N}$, which is plotted in
figure~\ref{fig:potential_pw}(a) for varying  $\sigma$ and for different
values of $N$. We can see that, for each $N$,  there is a neighbourhood up to
some finite value  $\sigma_c$  where it is vanishingly small, say
$K_N(\sigma)<\kappa$ with $\kappa$ being an arbitrary small value.  For
$\sigma$ outside this region, this coefficient rapidly transitions to the
value $1$, implying that the overdamped Brownian motion is no longer
inhibited by the multiscale fluctuations.  As can be seen in figure
\ref{fig:potential_pw}(a), increasing the number of scales $N$ moves this
transition point to higher $\sigma$. One can estimate such a transition point
by expanding  the function  $\cosh{(\frac{1}{\sigma})}=
1+\frac{1}{2}\frac{1}{\sigma^2}+\frac{1}{4!}\frac{1}{\sigma^4}+O(\sigma^{-6})$
so that
at the transition point $\sigma_c$ we have:
\begin{equation}
K_N(\sigma_c)\sim  \frac{1}{\left(1+\frac{1}{12}\frac{1}{\sigma_c^2}\right)^N}.
\end{equation}
We are interested in finding the value of $\sigma_c$ for which $K_N(\sigma_c)=\kappa$. By taking the logarithm of the above expression
we obtain at leading order:
\begin{equation}
\sigma_c(N) \sim \frac{1}{\sqrt{12\vert\log{\kappa}\vert}} N^{1/2},
\end{equation}
which is valid for sufficiently large values of $N$.

To further demonstrate the effect of the scales on the rate of diffusion of $X^\epsilon (t)$, we numerically simulate (\ref{e:gen-sde}) for the piecewise potential (\ref{eq:pw pot}) for different values of $N$.  For $\epsilon=0.1$ and $\sigma = 1$,  we approximate the solution of (\ref{e:gen-sde}) up to time $T = 100$, using an Euler-Maruyama discretisation with step-size varying between $10^{-7}$ and $10^{-8}$.  Given the resulting approximation $X_1, X_2, \ldots, X_M$, the diffusion coefficient was approximated by using the maximum likelihood estimator:
\begin{equation}
\label{e:eff_diff_estimator}
	\hat{D}(\lbrace X_1, \ldots, X_M\rbrace) = \frac{1}{\lfloor M/k\rfloor}\sum_{i=1}^{\lfloor M/k\rfloor} \frac{\left(X_{i + k} - X_i\right)^2}{2k\Delta t},
\end{equation}
where $k \in \mathbb{N}$ controls the subsampling time $\delta = k \Delta t$.
As noted in \cite{PavlSt06}, when inferring transport coefficients from
multiscale data, the subsampling rate must be chosen carefully to ensure that
the estimator converges to the diffusion coefficient of (\ref{e:gen-sde}) on
the $O(1)$ timescale (i.e. the effective diffusion coefficient).  Based on
short numerical experiments, $k = 10^{-1}/\Delta t$ was used.  The estimator
(\ref{e:eff_diff_estimator}) was then averaged over $100$ independent
realisations. In figure \ref{fig:potential_pw}(b) we show the average of this
estimator over $100$ realisations, as a function of time, for $N=1$, $N=2$
and $N=3$ scales respectively.  The error bars denote $95\%$ confidence
intervals.  The dashed lines denote the homogenized effective diffusion
coefficient $M(\sigma)$ predicted by  (\ref{e:pw M}).  We see good agreement
in each case,  although the discrepancy between the simulated diffusion
coefficient and $M(\sigma)$ increases as $N$ increases.  This discrepancy is
likely caused by discretisation error  due to the increase in stiffness for
larger values of $N$, as well as the fact that the small scale parameter
$\epsilon$ might not be sufficiently small to faithfully capture the
homogenized dynamics.


%
\section{Conclusions}
\label{sec:conclusions}

We have analysed the overdamped Langevin dynamics of a Brownian particle
moving in a multiscale potential. Using multiscale techniques we derived a
coarse-grained equation with a space-dependent diffusion tensor (i.e. with
multiplicative noise), driven by the system's free energy. The calculation of
the diffusion tensor requires the solution of a coupled-system of $N$ Poisson
equations. This system can be solved in one dimension and an explicit formula
(up to quadratures) for the diffusion coefficient can be obtained.

We demonstrated that the system can exhibit noise/multiscale-induced
transitions and these were analyzed in different types of multiscale
potentials. In the case of a double well potential with one non-separable
microscopic scale, it was shown that the multiscale structure can induce
hysteresis effects in the pitchfork bifurcation, something that was observed
to be enhanced as the number of microscopic scales was increased. For the
case of a tilted three-scale quartic potential we have shown that the
presence of the microstructure is able to change the bifurcation diagram such
that the resulting effect is that the stability of the zero solution can be
tuned by changing the noise intensity. The diffusion coefficient was
calculated analytically for a piecewise linear potential at all scales and
the transitions in the limit of infinitely many scales were studied.

The present works opens up several new avenues for research. First, the study
of noise/multiscale induced transitions in higher dimensions and the
construction of the corresponding bifurcation diagram would be a natural
extension. Furthermore, it would be interesting to study the effect of
inertia on the coarse-grained dynamics. Homogenization problems for the
underdamped Langevin dynamics~\cite{HP07} or, even more so, for the
generalized Langevin equation~\cite{OttobrePavliotis11} are technically more
challenging due to the hypoelliptic nature of the coresponding Fokker-Planck
operator.
%
In addition, the study of mean field limits for interacting multiscale
diffusions, in the sense of~\cite{Dawson1983} is a very challenging problem.
Finally, it would be also interesting to consider the effect of nonreversible
perturbations in the multiscale Brownian dynamics. Such a problem is relevant
for developing improved sampling techniques for multiscale
diffusions~\cite{LelievreNierPavliotis2013, DuncanLelievrePavliotis2016}. We
shall consider these and related issues in future studies.

\section*{Acknowledgements}
We are grateful to Prof. Yannis Kevrekidis for numerous stimulating
discussions, insightful comments and suggestions. We acknowledge financial
support by the Engineering and Physical Sciences Research Council of the UK
through Grants Nos. EP/H034587, EP/J009636, EP/K008595, EP/L020564,
EP/L024926, EP/L025159, EP/L027186 and EP/N005465 as well as European
Research Council through Advanced Grant No. 247031.
%
%
\section*{References}
\bibliographystyle{plain}
\input{mybib.bbl}

\appendix
\section{Brownian motion in an $N-$scale potential: Derivation of the homogenized equation}
\label{app:Reiterated}
Consider the following $\mathbb{R}^d$--valued overdamped Langevin diffusion process corresponding to the multiscale potential $V^\epsilon$:
\begin{equation}
dx^\epsilon_t = -\nabla V^\epsilon(x^\epsilon_t)\, dt + \sqrt{2\sigma}\, dW_t,
\end{equation}
where $W_t$ is a $d$-dimensional standard Brownian motion, and where the $N$--scale potential $V^\epsilon$ satisfies
\begin{equation}
V^\epsilon(x) = V(x, x/\epsilon, x/\epsilon^2, \ldots, x/\epsilon^N),
\end{equation}
for some smooth $V:\mathbb{R}^d\times \mathbb{T}^d\times\ldots\times\mathbb{T}^d \rightarrow \mathbb{R}$.  Given a smooth observable $f:\mathbb{R}^d\rightarrow \mathbb{R}$, the time evolution of the expectation $F^\epsilon(x, t) = \mathbb{E}_{x^\epsilon_0=x}\left[f(x^\epsilon(t))\right]$ satisfies the following backward Kolmogorov equation (BKE)
\begin{equation}
\partial_t F^\epsilon(x, t) = \mathcal{L}^\epsilon F(x, t),
\end{equation}
where the operator $\mathcal{L}^\epsilon$ is the infinitesimal generator  $\mathcal{L}^\epsilon$ is defined by
$$
	\mathcal{L}^\epsilon f(x) = -\nabla V^\epsilon(x)\cdot\nabla f(x) + \sigma \Delta f(x),	\quad f \in C^2_0(\mathbb{R}^d).
$$
We shall use reiterated homogenization to identify the behaviour of $F^\epsilon(x, t)$ in the limit as $\epsilon\rightarrow 0$.  We shall follow  the formal approach described in \cite[Section 3.7]{lions},  namely of ``freezing" the scales  $x, x/\epsilon, \ldots, x/\epsilon^{N-1}$ and studying the macroscopic effects of the $O(\epsilon^{-N})$ oscillations using classical periodic homogenization.  To this end, we shall formally assume that the variable $x/\epsilon^N$ is independent from the variables $x, x/\epsilon, \ldots, x/\epsilon^{N-1}$, writing $V^\epsilon(x) = V_N^\epsilon(x, x/\epsilon^N)$, so that
$$
	\nabla_x V^\epsilon(x) = \left(\nabla_x + \frac{1}{\epsilon^N}\nabla_z\right)V^\epsilon_N(x, z)\Big|_{z = x/\epsilon^N}.
$$
We shall look for solutions $F^\epsilon(x,t)$ of the form $F(x, x/\epsilon^N, t)$ where
\begin{equation}
\label{eq:ansatz}
	F(x, z, t) = F_0(x, z, t) + \epsilon F_1(x, z, t) + \epsilon^2 F_2(x, z, t) + \ldots.
\end{equation}
The Backward Kolmorogov equation can be rewritten as
\begin{eqnarray}
\label{eq:kbe_frozen}
	\partial_t F(x, z, t) = &-D_N V^\epsilon(x, z)D_N F(x, z, t) + \sigma D_N\cdot D_N F(x, z, t),
\end{eqnarray}
where $D_N = (\nabla_x + \epsilon^{-N}\nabla_z)$.  We now perform a standard homogenization procedure of the above PDE to obtain the effective dynamics in the limit of $\epsilon\rightarrow 0$.  We substitute this ansatz (\ref{eq:ansatz}) in (\ref{eq:kbe_frozen}) and consider the leading order terms of the expansion in powers of $\epsilon^{-1}$.  The $O(\epsilon^{-2N})$ can be written as
\begin{equation}
\label{eq:eps_m_2N}
	\nabla_z\cdot\left(e^{- V^\epsilon_N(x,z)/\sigma}\nabla_z F_0(x, z,t)\right)=0,\quad z \in \mathbb{T}^d
\end{equation}
Since for fixed $x$,  $e^{- V_N^\epsilon(x, z)/\sigma} > 0$ uniformly on $\mathbb{T}^d$, (\ref{eq:eps_m_2N}) implies that $F_0$ does not depend on the fast variable, i.e. $F_0(x, z, t) = F_0(x, t)$, $\forall (x, t) \in \mathbb{R}^d\times [0,\infty)$. The $O(\epsilon^{-N})$ equation is given by
\begin{equation}
\label{eq:eps_m_2N_p1}
	\nabla_z\cdot\left(e^{-V^\epsilon_N(x,z)/\sigma}\nabla_z F_1(x,z,t)\right) = -\nabla_z\cdot\left(e^{- V^\epsilon_N(x,z)/\sigma}\right)\partial_x F_0(x,z,t), \quad z \in \mathbb{T}^d.
\end{equation}
Let $\theta_N(x, z)$ be the vector valued solution of the following Poisson equation:
\begin{equation}
\label{eq:cellN}
	\nabla_z\cdot\left(e^{- V^\epsilon_N(x,z)/\sigma}\left(\nabla_z \theta_N(x,z) + I\right)\right)  = 0,	\quad z \in \mathbb{T}^d,
\end{equation}
where $(\nabla_z\theta_N)_{ij} = \partial_{z_i}\theta_{N,j}$, for $i,j=1,\ldots,d$. It is clear that $F_1(x,z,t) = \theta_N(x,z)\nabla_x F_0(x,t)$ satisfies (\ref{eq:eps_m_2N_p1}).  Finally, consider the $O(1)$ equation given by
\begin{eqnarray*}
\nabla_z\cdot\left(e^{-V^\epsilon_N(x,z)/\sigma}\nabla_z F_2(x,z,t)\right) = &-&\nabla_z\cdot\left(e^{- V^\epsilon_N(x,z)/\sigma}\nabla_x F_1(x,z,t)\right)  \\
	&-&\nabla_x\cdot\left(e^{- V^\epsilon_N(x,z)/\sigma}\nabla_z F_1(x,z,t)\right) \\
	&-&\nabla_x\cdot\left(e^{- V^\epsilon_N(x,z)/\sigma}\nabla_x F_0(x,t)\right)\\
	&-& \frac{e^{- V^\epsilon_N(x,z)/\sigma}}{\sigma} \partial_t F_0(x, t).
\end{eqnarray*}
A necessary and sufficient condition for $F_2$ to exist, is that the RHS has integral zero with respect to $e^{-V_N(x,z)}\,dz$, i.e.
\begin{eqnarray*}
	 Z_{N-1}(x)\partial_t F(x, t) &=& \sigma\nabla_x\cdot \left(\int e^{- V_N^\epsilon(x,z)/\sigma}\nabla_z F_1(x, z, t)\,dz\right)  \\
	 & &\quad + \sigma\nabla_x \left(\int e^{- V_N^\epsilon(x,z)/\sigma}\,dz \nabla_x F_0(x, t)\right) \\
	&=& \sigma\nabla_x\cdot \left(\int e^{- V_N^\epsilon(x,z)/\sigma} (\nabla_z \theta_N + I)\,dz\, \nabla_x F_0(x,t)\right) \\
	&=& \nabla_x\cdot \left(K_{N-1}(x)\, \nabla_x F_0(x,t)\right)
\end{eqnarray*}
where
$$Z_{N-1}(x) = \int e^{- V^\epsilon_N(x, z)/\sigma}\,dz,$$
and
$$K_{N-1}(x)= \sigma \int e^{- V_N^\epsilon(x,z)/\sigma} (\nabla_z \theta_N(x,z) + I)\,dz.$$
We now repeat the homogenization process, assuming that the $O(\epsilon^{-(N-1)})$ term is independent from the coarser scales, by reintroducing the small scales the above coarse grained PDE.  To this end writing
$$
	Z_{N-1}(x) = Z_{N-1}^{\epsilon}(x, x/\epsilon^{N-1}), \quad \mbox{ and }\quad 	K_{N-1}(x) = K_{N-1}^{\epsilon}(x, x/\epsilon^{N-1}),
$$
and $D_{N-1} = (\nabla_x + \frac{1}{\epsilon^{N-1}}\nabla_z)$
the KBE after coarse-graining the $O(\epsilon^{-N})$ fluctuations can be written as
$$
	\partial_t F^\epsilon(x, z, t) = \frac{\sigma}{Z^\epsilon_{N-1}(x, z)}D_{N-1}\cdot\left(K_{N}(x, z)D_{N-1} F^\epsilon(x, z, t)\right),
$$
where $D_{N-1} = (\nabla_X + \frac{1}{\epsilon^{N-1}}\nabla_z)$.  This can now be homogenized in an analogous manner.  Suppose now that this homogenization process has been repeated $k$ times so that the resulting coarse-grained PDE is given by
\begin{equation}
\label{eq:kbe_k}
\partial_t F^\epsilon(x, z, t) = \frac{\sigma}{Z^\epsilon_{N-k}(x, z)}D_{N-k}\cdot\left(K^\epsilon_{N-k}(x, z)D_{N-k} F^\epsilon(x, z, t)\right),
\end{equation}
where
$$
	Z^\epsilon_{N-k}(x, z) =\int\ldots\int e^{- V_{N-k}^\epsilon(x, z, x_{N-k+1}, \ldots, x_N)/\sigma}dx_N \ldots dx_{N-k+1}.	
$$
and
$$
K^\epsilon_{N-k}(x, z) = \sigma\int\cdots\int(I + \nabla_{x_N}\theta_N)\ldots (I + \nabla_{x_{N-k}}\theta_{N-k}) e^{-V^\epsilon_{N-k}(x, z, x_{N-k+1,\ldots, x_N})}\,dx_N\ldots, dx_{N-k+1}.
$$
and $D_{N-k} = \nabla_x +\frac{1}{\epsilon^{N-k}}\nabla_z$.  Once again, we look for solutions of $F^\epsilon(x, z, t)$ of the form
$$
	F^\epsilon(x, z, t) = F_0(x, z, t) + \epsilon F_1(x, z, t) + \epsilon^2 F_2(x, z , t) + \ldots.
$$
Substituting this ansatz in (\ref{eq:kbe_k}) we obtain leading order equation:
$$
	\nabla_{z}\cdot\left(K_{N-k}^\epsilon(x, z) \nabla_z F_0(x, z, t)\right) = 0,
$$
and since for fixed $x \in \mathbb{R}^d$,  $K^\epsilon_{N-k}(x,z) > 0$ over $\mathbb{T}^d$, it follows that $F_0(x, z, t) = F_0(x, t)$. The next leading order equation is given by
\begin{equation}
\label{eq:F1_k}
	\nabla_{z}\cdot\left(K_{N-k}^\epsilon(x, z) \nabla_z F_1(x, z, t)\right) = -\nabla_{z}\cdot\left(K_{N-k}^\epsilon(x, z) \nabla_x F_0(x, z, t)\right).
\end{equation}
Letting $\theta_{N-k}$ be the solution of the cell equation
\begin{equation}
\label{eq:cell_k}
	\nabla_{z}\cdot\left(K_{N-k}^\epsilon(x, z) \left(\nabla_z \theta_{N-k}(x, z) + I\right)\right) = 0,
\end{equation}
then choosing $F_1(x, z, t) = \theta_{N-k}(x,z)\nabla_{x}F_0(x, t)$ satisfies (\ref{eq:F1_k}).  The next equation in the expansion is then given by
\begin{eqnarray*}
	\nabla_{z}\cdot\left(K_{N-k}^\epsilon(x, z) \nabla_z F_2(x, z, t)\right) = &-&\nabla_{z}\cdot\left(K_{N-k}^\epsilon(x, z) \nabla_x F_1(x, z, t)\right) \\
	&-&\nabla_{x}\cdot\left(K_{N-k}^\epsilon(x, z) \nabla_z F_1(x, z, t)\right)\\
	&-&\nabla_{x}\cdot\left(K_{N-k}^\epsilon(x, z) \nabla_x F_0(x, t)\right)\\
	&-& Z^\epsilon_{N-k}(x,z)\partial_t F_0(x,t).
\end{eqnarray*}
A necessary and sufficient condition for $F_2$ to exist is that the RHS has integral zero with respect to $Z^{\epsilon}_{N-k}(x, z)\,dz$, i.e.
\begin{eqnarray*}
	 \int Z^{\epsilon}_{N-k}(x, z)\,dz \partial_t F_0(x,t) = \nabla_x\cdot\left(\int K_{N-k}^\epsilon(x,z)\left(\nabla_z \theta_{N-k}(x,z) + I\right) \nabla_x F_0(x,t)\right)
\end{eqnarray*}
Denote by
$$
Z_{N-k-1}^\epsilon(x) = \int Z^{\epsilon}_{N-k}(x, z)\,dz = \int\ldots\int e^{- V_{N-k}^\epsilon(x, x_{N-k}, x_{N-k+1}, \ldots, x_N)/\sigma}\,dx_{N-k}\, dx_{N-k+1}, \ldots, dx_N.
$$
We can then choose
 $$
	K_{N-k-1}^{\epsilon}(x) := \int K_{N-k}^\epsilon(x,z)\left(\nabla_z \theta_{N-k}(x,z) + I\right)\,dz,
$$
so that the PDE after coarse graining the $(N-k-1)^{th}$ scale becomes
$$
	\partial_t F_0(x,t) = \frac{1}{Z_{N-k-1}(x)}\nabla_x\cdot\left(K_{N-k-1}(x) \nabla_x F_0(x,t)\right).
$$
Following the above inductive scheme $N$ times, we obtain the following coarse grained PDE which is independent of $\epsilon$:
\begin{equation}
\label{eq:kbe_final}
	\partial_t {F}^0(x,t) = \frac{1}{Z(x)}\nabla_x\cdot\left(K(x)\nabla_x {F}^0(x,t)\right),
\end{equation}
where
$$
	Z(x) = \int\ldots \int e^{- V(x, x_1,\ldots, x_N)/\sigma}\,dx_1\ldots dx_{N},
$$
and
\begin{eqnarray*}
	K(x) &= \sigma\int\ldots\int \prod_{i=N}^{1} (1 + \nabla_{x_i}\theta_i(x, x_1,\ldots, x_i))e^{- V(x, x_1, \ldots, x_{N})/\sigma}\,dx_N\ldots dx_{1},
\end{eqnarray*}
where the correctors $\theta_1 ,\ldots, \theta_N$ are the solutions (known up to additive constants) of (\ref{eq:cell_k}).\\\\
We can observe that (\ref{eq:kbe_final}) corresponds to the BKE of a diffusion process $x^0$ described by the following SDE:
\begin{equation}\label{Eq:Homogenised}
dx^0_t = [- \mathcal{M}(x^0_t)\nabla_x\Psi(x^0_t)+ \nabla_x\cdot \mathcal{M}(x^0_t)]\, dt + \sqrt{2 \mathcal{M}(x^0_t)} dW_t,
\end{equation}
where we have defined:
\begin{equation}\label{Eq:M_0}
\mathcal{M}(x) =\frac{K(x)}{Z(x)}
\end{equation}
and
\begin{equation}
\Psi(x) = - \log{Z(x)},
\end{equation}%
One can moreover show that the matrix $\mathcal{M}(x)$ is symmetric positive definite, and therefore a matrix square root $\sqrt{\mathcal{M}(x)}$ is guaranteed to exist.
\\\\
This  result suggests that the process $x_t^\epsilon$ converges weakly to $x_t^0$ as $\epsilon\rightarrow 0$.  In \cite{DuncanPavl2016} this convergence is obtained rigorously, subject to assumptions on the range of the multiscale fluctuations arising from $V^\epsilon$.

\section{Calculation of the Effective Diffusion Coefficient in one dimension}
\label{app:1d-deff}

In general, one is not able to obtain explicit expressions for the coefficients of the coarse-grained SDE, and one typically must resort to computational methods to approximate $M(x)$, for example solving for $\theta_1,\ldots, \theta_N$ using a numerical PDE solver.  However, in the particular case when $d=1$, we can obtain closed-form solutions for the cell equations, from which the effective diffusion coefficient can be readily calculated.  Indeed, the cell equation for the corrector $\theta_N$ in one dimension is given by
$$
	\partial_{x_N}\left(e^{-V(x_0, x_1, \ldots, x_N)/\sigma}\left(\partial_{x_N}\theta_N + 1\right)\right) = 0,
$$
so that
$$
	\partial_{x_N}\theta_N(x_0,\ldots, x_N) + 1 = C(x_0, \ldots, x_{N-1})e^{V(x_0, \ldots, x_{N})/\sigma},
$$
where
$$
C(x_0, \ldots, x_{N-1}) =\left(\int_{\mathbb{T}}e^{V(x_0, \ldots, x_N)/\sigma}\,dx_N\right)^{-1}.
$$
The effective diffusion coefficient $\mathcal{K}_{N-1}$ obtained after homogenizing the $N^{th}$ scale is then given by
$$
	\mathcal{K}_{N-1}(x_0, \ldots, x_{N-1}) = \sigma\int_{\mathbb{T}} ( 1 + \partial_{x_N}\theta_{x_N}(x_0, \ldots, x_N))\,dx_N = \sigma\left(\int_{\mathbb{T}}e^{V(x_0, \ldots, x_N)/\sigma}\,dx_N\right)^{-1}.
$$
Proceeding inductively from $N$ to $1$, if we assume that $\mathcal{K}_{N-k}$ has the form
$$
	\mathcal{K}_{N-k}(x_0, \ldots, x_{N-k}) = \sigma\left(\int_{\mathbb{T}}\cdots\int_{\mathbb{T}} e^{V(x_0, \ldots, x_N)/\sigma}\,dx_N\ldots, dx_{N-k+1} \right)^{-1},
$$
then
$$
	1 + \partial_{x_{N-k}}\theta_{N-k}(x_0, \ldots, x_{N-k}) = C(x_0, \ldots, x_{N-k-1})\mathcal{K}^{-1}_{N-k}(x_0, \ldots, x_{N-k})
$$
where
\begin{align*}
C(x_0, \ldots, x_{N-k-1}) &= \left(\int \mathcal{K}_{N-k}(x_0, \ldots, x_{N-k})^{-1}\, dx_{N-k}\right)^{-1}
\end{align*}
so that
\begin{align*}
\mathcal{K}_{N-k-1}(x_0,\ldots, x_{N-k-1}) &=\int \mathcal{K}_{N-k}(x_0,\ldots, x_{N-k})\left(\theta_{x_{N-k}}(x_0,\ldots, x_{N-k}) + 1\right)\,dx_{N-k}\\
&= \sigma\left(\int_{\mathbb{T}}\cdots\int_{\mathbb{T}} e^{V(x_0, \ldots, x_N)/\sigma}\,dx_N\ldots, dx_{N-k} \right)^{-1} \\
 &= \sigma\left(\int_{\mathbb{T}}\cdots\int_{\mathbb{T}} e^{V(x_0, \ldots, x_N)/\sigma}\,dx_N\ldots dx_{N-k} \right)^{-1}.
\end{align*}
Continuing this procedure inductively, it follows that the effective diffusion coefficient $\mathcal{M}(x) = Z(x)^{-1}\mathcal{K}_1(x)$ can be written as
\begin{equation}
\label{eq:eff_diff_1d}
	\mathcal{M}(x) = \frac{\sigma}{Z(x)\hat{Z}(x)},
\end{equation}
where
$$
\hat{Z}(x) = \int\cdots\int e^{V(x, x_1,\ldots, x_N)/\sigma}\,dx_1\ldots dx_N.
$$
In the special case where the scales in the potential are completely separated, i.e. when
$$
V^\epsilon(x) = V_0(x) + V_1(x/\epsilon) + \ldots + V_N(x/\epsilon^N),
$$
for a smooth confining potential $V_0$ and smooth periodic functions $V_1,\ldots, V_N$, then one can see from (\ref{eq:eff_diff_1d}) that the effective diffusion coefficient $\mathcal{M}(x)$ tensorises into a product of the form
\begin{equation}
\label{eq:eff_diff_1d_separable}
\mathcal{M}(x) = \sigma \prod_{i=1}^{N} \left(\int_{\mathbb{T}} e^{-V_i(y_i)/\sigma}\,dy_i\int_{\mathbb{T}} e^{V_i(z_i)/\sigma}\,dz_i\right)^{-1}.
\end{equation}
The contribution of each scale to the potential satisfies
$$
	\int_{\mathbb{T}} e^{-V_i(u_i)/\sigma}\,du_i \int_{\mathbb{T}} e^{V_i(v_i)/\sigma}\,dv_i \geq \left(\int_{\mathbb{T}} e^{(V_i(u_i) - V_i(u_i))/\sigma}\,du_i\right)^2 = 1,
$$
by the Cauchy Schwartz inequality, with equality holding only when $V_i = 0$.
This implies that that adding increasingly fine scale fluctuations to
$V^\epsilon$ will always decrease the effective diffusion coefficient, as one
would expect.

\end{document}

%% file: mybib.bbl
\def\cprime{$'$} \def\cprime{$'$} \def\cprime{$'$} \def\cprime{$'$}
  \def\cprime{$'$} \def\cprime{$'$} \def\cprime{$'$} \def\cprime{$'$}
  \def\Rom#1{\uppercase\expandafter{\romannumeral #1}}\def\u#1{{\accent"15
  #1}}\def\Rom#1{\uppercase\expandafter{\romannumeral #1}}\def\u#1{{\accent"15
  #1}}\def\cprime{$'$} \def\cprime{$'$} \def\cprime{$'$} \def\cprime{$'$}
  \def\cprime{$'$} \def\cprime{$'$} \def\cprime{$'$}
  \def\polhk#1{\setbox0=\hbox{#1}{\ooalign{\hidewidth
  \lower1.5ex\hbox{`}\hidewidth\crcr\unhbox0}}} \def\cprime{$'$}
  \def\cprime{$'$} \def\cprime{$'$}